\documentclass[lettersize,journal]{IEEEtran}
\usepackage{amsmath,amsfonts}
\usepackage{algorithmic}
\usepackage{algorithm}
\usepackage{array}
\usepackage[caption=false,font=normalsize,labelfont=sf,textfont=sf]{subfig}
\usepackage{textcomp}
\usepackage{stfloats}
\usepackage{url}
\usepackage{verbatim}
\usepackage{graphicx}
\usepackage{cite}
\usepackage{textcomp}
\usepackage{amsthm}
\usepackage{siunitx}
\usepackage{longtable,tabularx}
\usepackage{amssymb}
\usepackage{bm}
\usepackage[mathscr]{euscript}
\usepackage{physics}
\usepackage{float}
\usepackage{mathtools}
\usepackage{array}
\usepackage{multirow}
\usepackage{caption}
\usepackage{titlesec}
\usepackage{hyperref}
\usepackage{xcolor}
\usepackage{authblk}
\allowdisplaybreaks

\newtheorem{prop}{Proposition}

\let\emptyset\varnothing

\begin{document}

\title{Robust Multi-Sensor Multi-Target Tracking Using Possibility Labeled Multi-Bernoulli Filter}

\author{
Han~Cai~\IEEEmembership{Member,~IEEE}, Chenbao Xue, Jeremie~Houssineau, Zhirun~Xue % <-this % stops a space
\thanks{H. Cai is with the School of Aerospace Engineering, Beijing Institute of Technology, Beijing 100081, China, e-mail: caihanspace@gmail.com.}% <-this % stops a space
\thanks{C. Xue is with the School of Aerospace Engineering, Beijing Institute of Technology, Beijing 100081, e-mail: chenbaospace@bit.edu.cn}
\thanks{J. Houssineau is with the Division of Mathematical Sciences, Nanyang Technological University, 50 Nanyang Ave, 639798, Singapore, e-mail: houssineau.j@gmail.com}% <-this % stops a space
\thanks{(Corresponding author: Jeremie Houssineau)}
\thanks{Z. Xue is with the School of Aerospace Engineering, Beijing Institute of Technology, Beijing 100081, China, e-mail: xuezhirun2020@163.com}
\thanks{Manuscript received XX XX, 2023.}
}

\markboth{Journal of \LaTeX\ Class Files,~Vol.~XX, No.~XX, XX~2023}%
{Shell \MakeLowercase{\textit{et al.}}: A Sample Article Using IEEEtran.cls for IEEE Journals}

% \IEEEpubid{0000--0000/00\$00.00~\copyright~2023 IEEE}

\maketitle

\begin{abstract}
With the increasing complexity of multiple target tracking scenes, a single sensor may not be able to effectively monitor a large number of targets. Therefore, it is imperative to extend the single-sensor technique to Multi-Sensor Multi-Target Tracking (MSMTT) for enhanced functionality. Typical MSMTT methods presume complete randomness of all uncertain components, and therefore effective solutions such as the random finite set filter and covariance intersection method have been derived to conduct the MSMTT task. However, the presence of epistemic uncertainty, arising from incomplete information, is often disregarded within the context of MSMTT. This paper develops an innovative possibility Labeled Multi-Bernoulli (LMB) Filter based on the labeled Uncertain Finite Set (UFS) theory. The LMB filter inherits the high robustness of the possibility generalized labeled multi-Bernoulli filter with simplified computational complexity. The fusion of LMB UFSs is derived and adapted to develop a robust MSMTT scheme. Simulation results corroborate the superior performance exhibited by the proposed approach in comparison to typical probabilistic methods.

\end{abstract}


\section{Introduction}

\IEEEPARstart{M}{ulti}-Sensor Multi-Target Tracking (MSMTT) is the core technology of autonomous sensor systems, which can achieve effective multiple targets tracking performance through integrating fusion from a group of cooperative sensors. This technology has been extensively applied across diverse domains including military reconnaissance and LEO mega constellation~\cite{Cai2022leo}. The scheme of MSMTT comprises two major components, i.e., multi-target filtering and multi-sensor fusion. The Random Finite Set (RFS) theory provides a comprehensive framework conducive to the design of advanced algorithms of multi-target tracking and the fusion of information from multiple sensors, which significantly influenced the development of MSMTT in the last decades~\cite{vo2019multi}. 

One primary challenge encountered in typical RFS filters lies in effectively handling diverse sources of uncertainty originating from the state transition process, observation, and the modeling of newborn targets. It is a common practice to assume complete knowledge and perfect characterization concerning all uncertain components inherent to an MSMTT system. This form of uncertainty represents an aleatory uncertainty, innately embedded within the system and impervious to elimination, thereby reflecting the inherent stochastic nature of the system. However, there exists another form of uncertainty, namely epistemic uncertainty, which is always disregarded in practical applications. Epistemic uncertainty arises from the absence of information pertaining to aspects such as the birth process of undetected targets.

Several multi-target tracking algorithms based on the concept of Uncertain Finite Set (UFS)~\cite{ristic2020target} have been developed recently to handle the epistemic uncertainty in multi-target systems, such as the possibility PHD filter~\cite{houssineau2021linear} and the possibility Generalized Labeled Multi-Bernoulli (GLMB) filter~\cite{cai2022glmb}. These methods represent both aleatoric and epistemic uncertainty properly using Outer Probability Measure (OPM)~\cite{cai2023possibilistic}, which achieves advanced multi-target tracking performance in the case of limited information~\cite{mahler2011cphd,punchihewa2018multiple}. Nevertheless, as the complexity of multiple target tracking scenarios continues to increase, relying solely on information from a single sensor may prove inadequate for effectively monitoring all desired targets. Therefore, it becomes imperative to extend the capabilities of single-sensor techniques to encompass a cooperative network of multiple sensors.

Several references have been proposed to conduct the problem of multi-sensor fusion in the framework of RFS. Generalized Covariance Intersection (GCI)~\cite{mahler2000optimal}, or namely Exponential Mixture Density (EMD)~\cite{uney2019fusion}, is a popular method that is combined with RFS filters for multi-sensor fusion.  Consensus algorithms can be designed using GCI by computing the weighted Kullback-Leibler Average of multi-target densities, which are represented by PHD~\cite{uney2019fusion}, CPHD~\cite{battistelli2013consensus}, and GLMB RFSs~\cite{fantacci2018robust}. However, the investigation of multi-sensor fusion methods within the framework of OPM theory and leveraging the most recent advancements in possibilistic multi-target tracking techniques for achieving enhanced performance in the MSMTT problems remains largely unexplored in the existing literature.

These concerns motivate science and technology research toward the development of advanced and robust multi-sensor information fusion methods whilst properly accounting for the partial knowledge of some aspects of the system. Ref.~\cite{Houssineau2023} presents an intuitive information fusion method based on the possibility Bernoulli filter~\cite{ristic2020target}, which can preserve the independence of information across the sensor network. Despite the widespread utilization of the Bernoulli filter in engineering since it accounts for target presence and absence in tracking, further adapting the methods to MSMTT may not be straightforward. Incorporating the newly developed possibility Generalized Labeled Multi-Bernoulli (GLMB) filter into multi-sensor fusion is a viable resolution, but this is a challenging task due to the intricate formulation of GLMB.

To resolve these issues, this paper proposes a possibilistic LMB filter leveraging the concept of OPMs. By utilizing OPMs, a flexible framework for handling various types of uncertain reasoning is established to manage situations where traditional probability theory may not be suitable due to incomplete or imprecise information. The derivation of the possibility LMB recursion is accomplished by approximating GLMB UFSs as LMB UFSs. Moreover, the concise formulation of the LMB UFS facilitates its application to the domain of multi-sensor fusion. On top of that, a robust MSMTT scheme is developed to properly account for the independence of information in the fusion. The effectiveness of the proposed possibilistic MSMTT method is validated by comparing it with conventional methods based on the probabilistic LMB filter and GCI fusion.

The major contributions of this paper, which go beyond previous references, can be summarized into three key aspects: 

(1) Possibility LMB filter: The paper introduces the possibility LMB filter, which is based on the concepts of labeled UFS and OPMs. The LMB possibility function is shown to be closed under prediction, and a suitable LMB approximation for the multi-target posterior is developed to preserve the LMB recursion. Leveraging these advancements, the possibility LMB filter demonstrates high robustness and simplified computational complexity in scenarios where information about some critical filtering parameters is incomplete.

(2) Multi-sensor fusion with LMB UFSs: The paper presents a closed-form solution of the multi-sensor fusion for multi-target distributions represented by LMB UFSs. This formulation represents a pioneering advancement for the fusion of labeled UFSs, offering a rigorous methodology for addressing the challenges of multi-sensor fusion in the presence of epistemic uncertainty. Furthermore, a label-matching strategy is proposed specifically for possibilistic LMB fusion, facilitating the integration of information acquired from sensors with varying Field-of-Views (FOVs).

(3) Flexible MSMTT framework for uncertain reasoning: The paper establishes a flexible framework that effectively manages various forms of uncertain reasoning by leveraging the possibility LMB filter and multi-sensor fusion formulations. The framework is applied to both centralized and distributed sensor networks, offering a simple and robust method that can be readily implemented in real-world scenarios. By incorporating OPMs, the framework enables robust handling of uncertainty, enhancing the accuracy and reliability of MSMTT systems in challenging scenarios.

The rest of the manuscript is structured as below. Sec.~\ref{Sec_GLMB} provides a brief introduction to the fundamental aspects of OPMs and the possibility GLMB UFS. Sec.~\ref{Sec_LMB} delineates the features of the proposed possibility LMB filter. Sec.~\ref{sec_fusion} detailed the fusion of LMB UFS and its Gaussian max-mixture implementation. Sec.~\ref{Sec_implementation} introduces the possibilistic MSMTT scheme applicable to centralized and distributed sensor networks. Simulation results and analysis are organized in Sec.~\ref{sec_Test}  and the conclusion can be found in the final section.

\section{Background} \label{Sec_GLMB}

The notations utilized in this paper are presented as follows. Single-target variables and multi-target variables are denoted by lower-case letters, e.g. $x$, $z$, $\ell$, and upper-case letters e.g., $X$, $Z$, $L$, respectively. In order to distinguish labeled variables, bold-type symbols are introduced to represent labeled states and distributions, e.g., $\bm{x}$, $\bm{X}$, and $\bm{\pi}$. All symbols regarding spaces are denoted by blackboard bold letters, such as the state space $X \in \mathbb{X}$, the measurement space $Z \in \mathbb{Z}$, and the label space $\ell \in \mathbb{L}$. For a given state space $\mathbb{X}$, its all finite subset is denoted by $\mathcal{F}(\mathbb{X})$, and $\mathcal{F}_n(\mathbb{X})$ represents the finite subsets of the space $\mathbb{X}$ with $n$ elements. 

\subsection{Outer Probability Measures} \label{Sec_OPM}

OPMs assign credibility to events in a state space $\bm{X}$, where credibility reflects the incomplete information we have about the occurrence of events. OPMs assume the measure of the entire state space $\bm{X}$ is equal to 1, meaning that the total credibility assigned to all possible events in $\bm{X}$ is 1. Conversely, a measure of 0 is assigned to an empty set, indicating that an event with no elements has no credibility. For a measurable subset $E$ of $\bm{X}$, the OPM $\bar{P}(E) \in [0,1]$ implies the credibility of an event $X\in E$, where $X$ represents an uncertain variable. It quantifies the degree to which we believe the uncertain variable $X$ belongs to the subset $E$. 

Unlike probability measures, OPMs satisfy the subadditivity property instead of the additivity assumption. In other words, the credibility of the union $A \cup B$ of two subsets $A$ and $B$ is less than or equal to the sum of their individual credibilities, i.e., $\bar{P}(A \cup B) \leq \bar{P}(A) +\bar{P}(B)$. OPMs provide an upper limit or an envelope for the probabilities assigned to events by probability measures. For a specific event $X \in E$, the upper bound of a probability measure $p(E)$ can be defined by the OPM $\bar{P}$ below~\cite{houssineau2018smoothing}
\begin{equation} \label{eq_opm_bound}
    1 - \bar{P}(\bm{X} \setminus E) \leq p(E) \leq \bar{P}(E).
\end{equation}
The above equation defines $\bar{P}(E)$ as the upper limit of the probability related to the event $X \in E$, and $\bar{P}(\bm{X} \setminus E)$ as the maximum probability of $X$ being in the complement of $E$.  Consequently, the lower bound of $p(E)$ can be expressed as $1 - \bar{P}(\bm{X} \setminus E)$.  

The most commonly used form of OPMs is defined by 
\begin{equation}
    \bar{P}(E) = \sup_{x\in E} f(x),
\end{equation}
where $f(\cdot)$ is a non-negative possibility function that describes the uncertain variable $x$. The supremum of any possibility functions equals one. Theoretically, the above definition is identical to the possibility distribution in possibility theory\footnote{possibility functions are often referred to as possibility distributions in the standard approach to possibility theory~\cite{dubois2015possibility}}~\cite{dubois2015possibility}.

The Gaussian Max-Mixture (GMM) possibility function is utilized in this paper to implement the possibility LMB fusion method. GMM can be expressed as a weighted maximum of Gaussian possibility functions $\bar{\mathcal{N}}$. For instance, a GMM with $L$ Gaussian components is given by~\cite{delande2023exploring}
\begin{equation}
    f(x) = \underset{1 \leq i\leq L}{\max}~ w^{(i)} \bar{\mathcal{N}} \Big( x; \mu^{(i)}, \Sigma^{(i)} \Big),
\end{equation}
where the weights subject to the constraints of $w^{(i)} \in [0, 1]$ and $\max_{1 \leq i\leq L} w^{(i)} =1$, and the formal definition of the Gaussian possibility function $\bar{\mathcal{N}}$ is defined by
\begin{equation}
    \bar{\mathcal{N}}(x; \mu, \Sigma)= \exp{-\frac{1}{2} (x-\mu)^T \Sigma^{-1} (x- \mu)},
\end{equation}
where $\mu \in \mathbb{R}^d$ denotes the possibilistic expected value, and the variance matrix $\Sigma$ is $d \times d$ positive definite. The probabilistic expected value is an average, whereas the possibilistic expected value represents the value that holds the least amount of information that would lead to its rejection. 

The formulation of GMM shares similarities with that of the probabilistic Gaussian Mixture (GM) model, where the PDF $p(x)$ is expressed as the summation of weighted components $p^{(l)}(x)$, with $l$ ranging from 1 to $L$, and subject to the constraint $\sum_{l=1}^L w^l=1$. Given a sufficient number of components, the GM model can approximate any continuously differentiable PDF within a specified non-zero error tolerance. Similarly, the GMM formulation is capable of approximating any smooth possibility function.

\subsection{Possibility Bayesian Filter}

Following the derivation presented in Ref.~\cite{houssineau2018smoothing}, it is demonstrated that a Bayesian filter based on credibility measures exhibits an equivalent predict-update recursion when all possibility functions involved are Gaussian. Supposing the information about the multi-target state $X_{k-1}$ at time $k-1$ is represented by a possibility function $\pi_{k-1}(X| Z_{1:k-1})$, and the transition dynamics governing the object's state evolution is modeled by a conditional possibility function $f_{k|k-1}(\cdot | X)$, then the predicted possibility function at time $k$ is computed using the subsequent equation
\begin{equation} \label{eq_opm_predict}
    \pi_{k}(X_k|Z_{1:k-1}) = \underset{X \in \mathcal{F}(\mathbb{X})}{\sup}~ \Big[ \pi_{k-1}(X| Z_{1:k-1}) f_{k|k-1}(X_k | X) \Big], 
\end{equation}
Compared to the conventional Chapman-Kolmogorov equation, which describes the evolution of a probability distribution over time, a similar relationship exists in Eq.~\eqref{eq_opm_predict}, but with a notable difference. Instead of integrating probability distributions over the entire state space, the maximum operation of the possibility function is employed in the credibility Bayesian filter.

In the update step, utilizing the principles of Bayesian inference, the prior possibility function $\pi_{k}(X_k|Z_{1:k-1})$ is revised by assimilating information derived from the observation $Z_k$ captured by all sensors. The resulting posterior possibility function at time $k$ is defined below:
\begin{equation} \label{eq_opm_update}
    \pi_k(X_k| Z_{1:k}) = \frac{ \pi_{k}(X_k| Z_{1:k-1}) ~l_k(Z_k |X_k)}{\underset{X \in \mathcal{F}(\mathbb{X}) }{\sup} ~  \pi_{k}(X | Z_{1:k-1}) ~l_k(Z_k |X)},
\end{equation}
where $l_k(Z_k |X_k)$ denotes the multi-target likelihood function, and the set supremum in the denominator can be computed using the following equation 
\begin{equation} \label{eq_sup_labeled}
    \sup_{{X} \in \mathbb{X}} {\pi}({X}) = \max_{0 \leq n \leq \infty} \Bigg[\sup_{ \{x_1, \cdots, x_n \} \in \mathbb{X}^n} \bm{\pi} \big( \{ x_1, \cdots, x_n \} \big) \Bigg]. 
\end{equation}
The update equation allows for incorporating new information and refining our knowledge about the system's state over time. It resembles Bayes' theorem in its structure, yet differs in the treatment of the integral operation, which is substituted with a maximum operation. Moreover, the likelihood function $l(Z_k|Z)$, representing the conditional possibility of the observation $Z_k$ given the state $Z$, is employed in lieu of a traditional probability distribution.

By iteratively updating the prior possibility function based on new observations, the credibility Bayesian filter provides an effective framework for state estimation and tracking in the presence of epistemic uncertainty. 

\subsection{GLMB UFS} \label{sec_glmb_UFS}

An Uncertain Finite Set (UFS), denoted as $X \in \mathcal{F}(\mathbb{X})$, refers to an uncertain variable that takes on finite set values, which is the foundation for deriving the possibility Bernoulli filter for single-target tracking~\cite{ristic2020target}. In the context of multi-target tracking, UFS is further enhanced by upgrading it to a labeled version, as described in Ref.~\cite{cai2022glmb}. This is achieved by incorporating target identity information into the framework. Specifically, each target in the labeled UFS is assigned a unique label $\ell=(k, j) \in \mathbb{L}$, where $k$ denotes the time of birth and $j$ represents the order of targets born at time $k$. The label of $(x, \ell)$ can be obtained by applying the projection function $\mathcal{L}((x, \ell)) = \ell$, and $\mathcal{L}: \mathbb{X}\times \mathbb{L} \rightarrow \mathbb{L}$. The labeled UFS $\bm{X} \in \mathcal{F}(\mathbb{X} \times \mathbb{L})$ can be defined as a UFS that combines the state space $\mathbb{X}$ with a discrete label space $\mathbb{L}$. For a realization $\bm{X}$ of labeled UFS on the space $\mathbb{X} \times \mathbb{L}$, each element in this set has a unique label. This means that the cardinality $|\mathcal{L}(\bm{X})|$ of the label set is equal to the cardinality $|\bm{X}|$ of the UFS itself.  

This labeling mechanism allows for the identification and tracking of individual targets within the UFS, enabling more precise and accurate multi-target tracking. In Ref.~\cite{cai2022glmb}, a set of labeled UFSs was introduced as the foundation for developing the possibility GLMB filter. This section provides a brief introduction to the pertinent ones used in this paper.

\subsubsection{Generalized Labeled Multi-Bernoulli UFS}

The most representative labeled UFS is the GLMB UFS $\bm{X}$ defined on $\mathbb{X} \times \mathbb{L}$. Its possibility function is expressed by the following equation~\cite{cai2022glmb}
\begin{equation} \label{eq_GLMB}
    \bm{\pi}(\bm{X}) = \Delta(\bm{X}) \max_{o \in \mathbb{O}} w^{(o)}(\mathcal{L}(\bm{X})) \Big[ f^{(o)} \Big]^{\bm{X}},
\end{equation}
where $\Delta(\bm{X}) = \delta_{|\bm{X}|}(|\mathcal{L}(\bm{X})|)$ denotes a distinct label indicator, which takes the value $\Delta(\bm{X}) = 1$ iff the number of elements in $\bm{X}$ is the same as the number of distinct labels in $\mathcal{L}(\bm{X})$ and $\Delta(\bm{X})=0$ otherwise; $\mathbb{O}$ indicates a discrete index space; the exponential denotes a product of possibility functions for single target state, i.e., $[f^{(o)}]^{\bm{X}}=\prod_{\bm{x} \in \bm{X}} {f^{(o)}(\bm{x})}$. A possibility function $f_c(n)= \max_{o\in\mathbb{O}} \max_{L\in \mathcal{F}_n(\mathbb{J})} w^{(o)}(J)$ is introduced to describe the cardinality of $\bm{X}$ equals to $n$. The presence function of an unlabeled GLMB UFS represents the possibility that there is at least one target with a given state, which is defined as 
\begin{equation} \label{eq_GLMB_v}
    \bar{v}(x) = \max_{o\in \mathbb{O}} \max_{\ell \in \mathbb{L}} \Big[ f^{(o)}(x,\ell) \max_{J\subseteq \mathbb{L}} 1_L(\ell) w^{(o)}(J) \Big], 
\end{equation}
which subject to the normalization $\sup_{x\in\mathbb{X}}=1$.

\subsubsection{$\delta$-Generalized Labeled Multi-Bernoulli UFS}

To develop a computationally efficient possibilistic multi-target tracker, a specialized variant of the GLMB UFS is introduced, known as the $\delta$-GLMB UFS. This variant incorporates a specific possibility function, which is defined as follows~\cite{cai2022glmb} 
\begin{equation} \label{eq_dGLMB}
    \bm{\pi}(\bm{X}) = \Delta(\bm{X}) \max_{(I,\xi) \in \mathcal{F}(\mathbb{L})\times\Xi} \delta_I(\mathcal{L}(\bm{X})) w^{(I,\xi)} \Big[ f^{(\xi)} \Big]^{\bm{X}},
\end{equation}
where the discrete space $\Xi$ denotes the history of association maps between targets and measurements. Eq.~\eqref{eq_dGLMB} is related to Eq.~\eqref{eq_GLMB} through the following transformation
\begin{align*}
     \mathcal{F}(\mathbb{L})\times\Xi = \mathbb{O},  ~ \delta_I(J) w^{(I,\xi)} = w^{(o)}(J), ~f^{(\xi)} = f^{(o)}. 
\end{align*}
The presence function of a $\delta$-GLMB UFS can be seen as an adaption of Eq.~\eqref{eq_GLMB_v}, which is given by  
\begin{equation} \label{eq_dGLMB_v}
\begin{split}
    \bar{v}(x) & = \max_{(I,\xi)\in \mathcal{F}(\mathbb{L})\times\Xi} \max_{\ell \in \mathbb{L}} \Big[ f^{(\xi)}(x,\ell) \max_{L\subseteq \mathbb{L}} 1_J(\ell) w^{(I,\xi)} \delta_I(J) \Big].
\end{split}
\end{equation}

\subsubsection{Labeled Multi-Bernoulli UFS} 

Another family member of the GLMB UFS is the LMB UFS. It differs from the $\delta$-GLMB UFS in that only one hypothesis is in the index space $\mathbb{O}$ in LMB. Furthermore, the LMB UFS can be seen as an extension of the multi-Bernoulli UFS by incorporating labels to nonempty Bernoulli components.  The possibility function of an LMB UFS is given by~\cite{cai2022glmb}
\begin{equation}
    \bm{\pi}(\bm{X}) = \Delta(\bm{X}) w(\mathcal{L}(\bm{X})) f^{\bm{X}}, 
\end{equation}
where the weight $w(\cdot)$ for a set $L$ of labels has a maximum value of one on the label space $\mathbb{L}$, and 
\begin{align}
    w(L) = \prod_{i \in \mathbb{L}} \tau^{(i)} \prod_{\ell \in L} 1_{\mathbb{L}}(\ell)\frac{\gamma^{(\ell)}}{ \tau^{(\ell)}}.  \label{eq_lmb_w}
\end{align}
The parameters $\tau$ and $\gamma$ represent the possibility of non-existence and existence, respectively. For a Bernoulli component labeled by $\ell$,  $\tau$ and $\gamma$ bound the probability of existence as below 
\begin{equation}
    r^{(\ell)} \in[1-\tau^{(\ell)}, ~ \gamma^{(\ell)}].
\end{equation}
An LMB UFS can be represented by a parameter set $\{ (\tau^{(\ell)}, \gamma^{(\ell)}, f^{(\ell)}) \}_{\ell \in \mathbb{L}}$. 

The LMB UFS can be simplified to a multi-Bernoulli UFS by disregarding all labeling information. Furthermore, the maximum value of the LMB UFS, which represents the most likely hypothesis, is equal to one. This result also holds true for the unlabeled version.

\subsection{Multi-Target Model} \label{Sec_Model}

The process of multi-target prediction can be decomposed into parallel evolution of each target considering potential birth and death behavior. Whether a target survives to the current time depends on a possibility $\lambda_s(x, \ell)$. Should the target persist, the labeled state will be updated to $(x_+, \ell)$ according to a transition possibility function $f(x_+|x,\ell)$. Alternatively, it may disappear with a possibility $\lambda_d(x, \ell)$. The state of multiple survival targets can be modeled as an LMB UFS denoted by $\{(\lambda_d(\bm{x}), \lambda_s(\bm{x}),f_s(\cdot|\bm{x})): \bm{x} \in \bm{X}\}$. Meanwhile, the possibilities of new targets born and remaining undiscovered are denoted by $\gamma_b$ and $\tau_b$, respectively. The state of multiple birth targets can also be represented by an LMB UFS  $\{(\tau_b^{(\ell)}, \gamma_b^{(\ell)}, f_b^{(\ell)}) \}_{\ell=1}^\mathbb{B}$. 

The multi-target state $\bm{X}_+$ at the next time is the combination of survival targets $\bm{X}_S$ and birth targets $\bm{X}_B$. Assuming $\bm{X}_S$ and $\bm{X}_B$ are independent of each other, the multi-target transition kernel is given by~\cite{cai2022glmb}
\begin{equation} \label{eq_transition}
    \bm{\phi}(\bm{X}_+ | \bm{X}) = \bm{\phi}_s(\bm{X}_S|\bm{X}) ~\bm{\phi}_b(\bm{X}_B),
\end{equation} 
where $\bm{\phi}_s$ and $\bm{\phi}_b$ are possibility functions of survival and birth objects. 

During the observation process, a single target state $\bm{x}\in \bm{X}$ can be detected according to the possibility of successful detection $d_s(\bm{x})$. The successful detection results in an observation $z \in W$, which is described by a likelihood function $l(z|\bm{x})$. Conversely, a target state $\bm{x}$ can also lead to a missed detection $\emptyset$ due to the possibility of detection failure $d_f(\bm{x})$. The functions $d_s(\bm{x})$ and $d_f(\bm{x})$ are introduced to define the range within which the actual detection probability $p_d$ can vary, i.e.,
\begin{equation}
    p_d(\bm{x}) \in [1-d_f(\bm{x}), ~ d_s(\bm{x})].
\end{equation}
The set of detection $W$ can be represented as a multi-Bernoulli UFS $\big\{(d_f(\bm{x}), d_s(\bm{x}), l(\cdot|\bm{x}) ) \big\}_{\bm{x}\in \bm{X}}(W)$. Moreover, a separate UFS $Y$ is employed to model the set of false observations. The possibility function $\pi_F$ associated with this UFS $Y$ is defined by the following expression~\cite{cai2022glmb}
\begin{equation}
    \pi_F(Y) = \prod_{z \in Y} \kappa(z) = \kappa^Y. 
\end{equation}
The function $\kappa(z)$ represents the possibility that an observation $z$ corresponds to a false alarm. By considering the possibility functions associated with the measurements set $W$ and the false alarms $Y=Z-W$, the multi-target likelihood function is given by
\begin{equation} \label{eq_meas}
     l(Z|\bm{X}) = \max_{W \subseteq Z} \Big[ \pi_d(W|\bm{X}) \pi_F(Z-W) \Big],
\end{equation}
where $\pi_d(W|\bm{X})$ is the possibility function of $W$.

%=================================================================
\section{Possibility Labeled Multi-Bernoulli Filter} \label{Sec_LMB}

In this section, the prediction and update of the possibility LMB filter are detailed in Sec.~\ref{sec_lmb_predict} and Sec.~\ref{sec_lmb_update}, respectively. The implementation of the joint prediction and update scheme of the possibility LMB filter is given in Sec.~\ref{Sec_joint}.

\subsection{LMB Prediction} \label{sec_lmb_predict}

In the context of labeled UFS, the multi-target posterior and the birth distribution can both be described by LMB UFS, and their possibility functions are defined as follows
\begin{align}
    \bm{\pi}(\bm{X}) &= \Delta(\bm{X}) w(\mathcal{L}(\bm{X})) f^{\bm{X}}, \\
    \bm{\pi}_b(\bm{X}) &= \Delta(\bm{X}) w_b(\mathcal{L}(\bm{X})) \big[f_b\big]^{\bm{X}},
\end{align}
where
\begin{align}
    w(L) & = \prod_{i \in \mathbb{L}} \tau^{(i)} \prod_{\ell \in L} 1_{\mathbb{L}}(\ell) \frac{\gamma^{(\ell)}}{ \tau^{(\ell)}}, \label{eq_w_LMB} \\
    w_b(L) & = \prod_{i \in \mathbb{B}} \tau_b^{(i)} \prod_{\ell \in L} 1_{\mathbb{B}}(\ell) \frac{\gamma_b^{(\ell)}}{ \tau_b^{(\ell)}}. 
\end{align}
Since LMB UFS is a special form of GLMB UFS, following the derivation in Ref.~\cite{cai2022glmb}, the predicted LMB can be expressed as a GLMB defined in the state space $\mathbb{X}$ and label space $\mathbb{L}_+= \mathbb{B} \cup \mathbb{L}$, i.e., 
\begin{equation}
    \bm{\pi}_+(\bm{X}_+) = \Delta(\bm{X}_+) w_+(\mathcal{L}(\bm{X}_+)) \big[ f_+\big]^{\bm{X}_+}, 
\end{equation}
where
\begin{align}
    w_+(I_+) & = w_b(\mathbb{B} \cap I_+) w_s(\mathbb{L}\cap I_+ ), \\
    w_s(L) & = \big[\eta_s\big]^L \max_{I \in \mathbb{L}} 1_I(L)  w(I) \big[\eta_d\big]^{I-L}\label{eq_wS},\\
    \eta_s(\ell) & = \sup_{x_+ \in \mathbb{X}} \sup_{x \in \mathbb{X}} \Big[ \lambda_s(x,\ell)f(x_+|x,\ell) f(x,\ell) \Big], \\
    \eta_d(\ell) &= \sup_{x \in \mathbb{X}} \lambda_d(x,\ell)f(x,\ell), \\
    f_+(x_+,\ell) & = 1_{\mathbb{L}}(\ell)f_s(x_+,\ell) +(1-1_{\mathbb{L}}(\ell))f_b(x_+,\ell), \\
    f_s(x_+,\ell) & = \frac{\sup_{x\in \mathbb{X}} [\lambda_s(x,\ell)f(x_+|x,\ell) f(x,\ell)] }{\eta_s(\ell)},
\end{align}
$\eta_s(\ell)$ and $\eta_d(\ell)$ are the survival possibility and non-survival possibility of track $\ell$, respectively, and they subject to the constraint that $\max\{\eta_s(\ell), \eta_d(\ell)\}=1$. In the case of survival LMB, the weight $w_s(\cdot)$ may not be the same as in the standard form of LMB weight \eqref{eq_w_LMB} due to the maximum term over $\mathbb{L}$. 
 
The formulations of the possibility LMB prediction are presented in the subsequent proposition, while an elaborate derivation is provided in Appendix~\ref{App_LMB_predict}. 
 
\begin{prop} \label{prop_predict}
    The multi-target posterior at the current time is an LMB UFS with the parameter set $\bm{\pi}=\{(\tau^{(\ell)}, \gamma^{(\ell)}, f^{(\ell)}) \}_{\ell\in\mathbb{L}}$ defined in the state space $\mathbb{X}$ and label space $\mathbb{L}$. The multi-target birth model is also an LMB described by the parameter set $\bm{\pi}_b=\{(\tau_b^{(\ell)}, \gamma_b^{(\ell)}, f_b^{(\ell)}) \}_{\ell\in\mathbb{B}}$ on $\mathbb{X}$ and the label space $\mathbb{B}$. Then, the multi-target prediction is an LMB UFS defined in the state space $\mathbb{X}$ and label space $\mathbb{L}_+= \mathbb{B} \cup \mathbb{L}$, given by
    \begin{equation}
        \bm{\pi}= \Big\{ \big( \tau_s^{(\ell)}, \gamma_s^{(\ell)}, f_s^{(\ell)} \big) \Big\}_{\ell\in\mathbb{L}} \bigcup \Big\{ \big( \tau_b^{(\ell)}, \gamma_b^{(\ell)}, f_b^{(\ell)} \big) \Big\}_{\ell\in\mathbb{B}},
    \end{equation}
    where 
    \begin{align}
        \tau_s^{(\ell)} & = \max \{ \tau^{(\ell)}, \eta_d(\ell) \gamma^{(\ell)} \}, \label{eq_presence}\\
        \gamma_s^{(\ell)} & = \eta_s(\ell) \gamma^{(\ell)} , \label{eq_absence}\\
        f_s^{(\ell)}(x_+) &= \frac{1}{\eta_s(\ell)} \sup_{x\in \mathbb{X}} \big[ \lambda_s(x,\ell)f(x_+|x,\ell) f(x,\ell) \big].
    \end{align}
\end{prop}

For any track $\ell$, the maximum of its prior possibility of non-existence or existence equals one, i.e.
\begin{equation}
\begin{split}
    \max\{ \tau_s^{(\ell)}, ~ \gamma_s^{(\ell)}\} &= \max\{ \max\{ \tau^{(\ell)}, ~ \eta_d(\ell) \gamma^{(\ell)} \}, ~ \eta_s(\ell) \gamma^{(\ell)} \} \\
    &= \max\{ \tau^{(\ell)}, ~\gamma^{(\ell)} \max\{\eta_d(\ell), ~\eta_s(\ell)\}\}\\
    &= \max\{ \tau^{(\ell)}, ~\gamma^{(\ell)}\}\\
    &= 1.
\end{split}
\end{equation}  
The above equation can be easily validated since $\max\{\tau^{(\ell)}, \gamma^{(\ell)}\}=1$ and  $\max\{\eta_s(\ell), \eta_d(\ell)\}=1$.

\subsection{LMB Update} \label{sec_lmb_update}

Note that the updated multi-target prior LMB does not yield a closed-form solution. Inspired by Ref.~\cite{reuter2014labeled}, we transform the prior LMB to the $\delta$-GLMB form for measurement update, and the multi-target posterior is then approximated by an LMB UFS for recursive estimation. The prior $\delta$-GLMB has the following form 
\begin{equation}
    \bm{\pi}(\bm{X}) = \Delta(\bm{X}) \max_{I_+ \in \mathcal{F}(\mathbb{L}_+)} w_+^{(I_+)} \delta_{I_+}(\mathcal{L}(\bm{X})) \big[f_+ \big]^{X},
\end{equation} 
where the weight $w_+^{(I_+)}$ and possibility function $f_+$ are given by 
\begin{align}
    w_+^{(I_+)} & = [\gamma_s]^{\mathbb{L} \cap I_+} [\tau_s]^{\mathbb{L} - I_+} [\gamma_b]^{\mathbb{B} \cap I_+} [\tau_b]^{\mathbb{B} - I_+}, \\
    f_+(x,\ell) & = 1_{\mathbb{L}}(\ell) f_s(x,\ell) + 1_{\mathbb{B}}(\ell) f_b(x,\ell). 
\end{align}

The posterior $\delta$-GLMB is computed following the update step of the possibilistic $\delta$-GLMB filter given in Ref.~\cite{cai2022glmb}, i.e., 
\begin{equation} \label{eq_postlmb}
\begin{split}
    \bm{\pi}(\bm{X}|Z) = \Delta(\bm{X}) & \max_{(I_+, \theta) \in \mathcal{F}(\mathbb{L}_+) \times \Theta_+} w_Z^{(I_+, \theta)} \\
    & \times \delta_{I_+}(\mathcal{L}(\bm{X})) \big[f_Z^{(\theta)} \big]^{X}, 
\end{split}
\end{equation}
where 
\begin{align}
    w^{(I_+,\theta)}_Z & = \delta_{\theta^{-1}(\{0:|Z|\})}(I_+) w_+^{(I_+)} \Big[\eta_Z^{(\theta)}\Big]^{I_+} \\
    f_Z^{(\theta)}(x,\ell) &= \frac{f_+(x,\ell) \psi_Z(x,\ell;\theta)}{\eta_Z^{(\theta)}(\ell)} \\
    \eta_Z^{(\theta)}(\ell) &= \sup_{x\in\mathbb{X}} f_+(x,\ell) \psi_Z(x,\ell;\theta) \\
    \psi_Z(x,\ell;\theta) & =
    \begin{cases}
      \frac{d_s(x,\ell)l(z_{\theta(\ell)}|x,\ell)}{\kappa(z_{\theta(\ell)})} & \text{if} ~ \theta(\ell) \neq 0, \\
      d_f(x,\ell) & \text{if} ~ \theta(\ell)=0.
    \end{cases}   
\end{align}

The $\delta$-GLMB formulation of the posterior multi-state state can then be transformed back to an LMB representation, and the parameters defining this transformation are defined in the following proposition. A comprehensive derivation of this conversion can be found in Appendix~\ref{App_LMB_update}. 

\begin{prop} \label{prop_update}
Given the predicted multi-target prior as an LMB UFS defined in the state space $\mathbb{X}$ and label space $\mathbb{L}$, and described by the parameter set $\bm{\pi}_+=\{(\tau_{+}^{(\ell)}, \gamma_{+}^{(\ell)}, f_+^{(\ell)}) \}_{\ell\in\mathbb{L}_+}$, the LMB approximation $\bm{\pi}=\{(\tau^{(\ell)}, \gamma^{(\ell)}, f^{(\ell)}) \}_{\ell\in\mathbb{L}_+}$ of the updated multi-target posterior is given by
\begin{align}
    \tau^{(\ell)} & = \max_{(I_+,\theta)\in\mathcal{F}(\mathbb{L}_+)\times\Theta_+} (1- 1_{I_+}(\ell)) w^{(I_+,\theta)}_Z \\
    \gamma^{(\ell)} & = \max_{(I_+,\theta)\in\mathcal{F}(\mathbb{L}_+)\times\Theta_+} 1_{I_+}(\ell) w^{(I_+,\theta)}_Z \\
    f^{(\ell)}(x) & = \frac{1}{\gamma^{(\ell)}} \max_{(I_+,\theta)\in\mathcal{F}(\mathbb{L}_+)\times\Theta_+} \notag \\
    & \qquad \times 1_{I_+}(\ell) w^{(I_+,\theta)}_Z f_Z^{(\theta)}(x,\ell). 
\end{align}

\end{prop}

For any track $\ell$, the maximum of its posterior possibility of non-existence or existence equals one, i.e., $\max\{\tau^{(\ell)}, \gamma^{(\ell)}\}=1$. Assume the posterior hypothesis with maximum weight is denoted as $(I^*,\theta^*)$, where
\begin{equation}
    w^{(I^*,\theta^*)}(Z)=\max_{(I,\theta)\in\mathcal{F}(\mathbb{L}_+)\times\Theta} w^{(I,\theta)}(Z)=1. 
\end{equation}
If a track $\ell\in I^*$, then its possibility of existence equals one, 
\begin{equation}
    \gamma^{(\ell)} = \max_{(I,\theta)\in\mathcal{F}(\mathbb{L}_+)\times\Theta} w^{(I,\theta)}(Z) 1_{I}(\ell)= w^{(I^*,\theta^*)}(Z)=1. 
\end{equation}
If a track $\ell\not\in I^*$, then its possibility of non-existence equals one, 
\begin{equation}
    \tau^{(\ell)} = \max_{(I,\theta)\in\mathcal{F}(\mathbb{L}_+)\times\Theta} w^{(I,\theta)}(Z) (1- 1_{I}(\ell))= w^{(I^*,\theta^*)}(Z)=1. 
\end{equation}
Therefore, $\max\{\tau^{(\ell)}, \gamma^{(\ell)}\}=1$ holds true for any tracks.

\subsection{Joint Prediction and Update} \label{Sec_joint}

The joint prediction and update implementation of the possibility $\delta$-GLMB filter has been developed in Ref.~\cite{cai2022glmb}. The $\delta$-GLMB possibility function at the current time can be obtained as a function of the $\delta$-GLMB possibility function at the last time as below
\begin{equation} \label{eq_joint_glmb}
\begin{split}
    \bm{\pi} (Z|\bm{X}) \propto \Delta (\bm{X}) \max_{(I, \xi, I_+, \theta)} & \omega^{(I, \xi)} \omega_{Z}^{(I, \xi, I_+, \theta)} \\
    & \times \delta_{I_+}(\mathcal{L}(\bm{X})) \Big[ f^{(\xi,\theta)}_Z \Big]^{\bm{X}},
\end{split}
\end{equation}
where 
\begin{align}
    \omega_{Z}^{(I, \xi, I_+, \theta)} & = 1_{\Theta(I_+)}(\theta) [\tau_b]^{\mathbb{B}-I_+} [\gamma_b]^{\mathbb{B} \cap I_+} \notag\\
    & \quad \times \Big[\eta_s^{(\xi)} \Big]^{I\cap I_+} \Big[\eta_d^{(\xi)} \Big]^{I-I_+} \Big[ \eta_Z^{(\xi,\theta)} \Big]^{I_+} \\
    \eta_s^{(\xi)}(\ell) & = \sup_{x \in \mathbb{X}} \sup_{x' \in \mathbb{X}} \Big[ \lambda_s(x',\ell)f(x|x',\ell) f^{(\xi)}(x',\ell) \Big] \\
    \eta_d^{(\xi)}(\ell) &= \sup_{x \in \mathbb{X}} \lambda_d(x,\ell)f^{(\xi)}(x,\ell) \\
    \eta_Z^{(\xi,\theta)}(\ell) &= \underset{x\in\mathbb{X}}{\sup} ~ f^{(\xi)}(x,\ell) \psi_Z(x,\ell;\theta) \\
    f^{(\xi,\theta)}_Z(x, \ell) & = \frac{f_+^{(\xi)}(x,\ell) \psi_Z(x,\ell;\theta)}{\eta_Z^{(\xi,\theta)}(\ell)} \\
    f_+^{(\xi)}(x,\ell) & = 1_{\mathbb{L}}(\ell)f_s^{(\xi)}(x,\ell) +(1-1_{\mathbb{L}}(\ell))f_b(x,\ell) \\
    f_s^{(\xi)}(x,\ell) & = \frac{\sup_{x'\in \mathbb{X}} [\lambda_s(x',\ell) f(x|x',\ell) f^{(\xi)}(x',\ell)] }{\eta_s^{(\xi)}(\ell)}.
\end{align}
Since the weight $\omega_{Z}^{(I, \xi, I_+, \theta)}$ and $w^{(I_+,\theta)}_Z$ contain similar terms, it is straightforward to derive the joint prediction and update implementation through generating posterior hypotheses directly from the prior LMB. The major difference is that $w^{(I_+,\theta)}_Z$ consists of the predicted possibility of existence $\gamma_s$ and non-existence $\tau_s$, while $\omega_{Z}^{(I, \xi, I_+, \theta)}$ is established based on the posterior possibility of existence $\eta_s^{(\xi)}$ and non-existence $\eta_d^{(\xi)}$. In addition, the notion of association history $\xi$ is omitted from $w^{(I_+,\theta)}_Z$ since the $\delta$-GLMB is approximated by the prior LMB. Hence, in the joint LMB filter, all the hypotheses in the posterior $\delta$-GLMB can be seen as the descendants of a single prior hypothesis, which consists of all the LMB components. The update step of the LMB filter requires converting the prior LMB to $\delta$-GLMB, this conversion may yield a significant computational load especially when the number of Bernoulli components is large. Such a conversion is avoided in the joint prediction and update procedure.

%=================================================================

\section{Fusion of LMB UFS} \label{sec_fusion}

The integration of information from multiple sensors presents significant challenges, particularly in handling local multi-target data with epistemic uncertainties. This section addresses this challenge by proposing a fusion formula specifically tailored to the developed LMB UFS framework. This fusion formula serves as the foundation for designing an MSMTT algorithm for both centralized and distributed sensor networks. 

To elaborate on the fusion of LMB UFSs, Section~\ref{sec_lmbfusion} provides an in-depth explanation, while the derivation of the primary solution is outlined in Appendix~\ref{App_fusion}. Furthermore, Section~\ref{Sec_label} addresses the label-matching problem within LMB fusion.

\subsection{Fusion of LMB UFS} \label{sec_lmbfusion}

In this section, the proposed approach aims to establish a comprehensive understanding of the solution involved in the fusion of LMB UFSs, facilitating the design and implementation of robust MSMTT algorithms in diverse sensor network scenarios. 

\begin{prop} \label{prop_fusion}
Given a set of $N$ LMB UFSs $\bm{\pi}_i$ with parameter set $\{(\tau_i^{(\ell)}, \gamma_i^{(\ell)}, f_i^{(\ell)}) \}_{\ell \in \mathbb{L}}$, $i=1,\cdots, N$ sharing the same label space, the fusion of these LMB UFSs is still an LMB UFS $\Tilde{\bm{\pi}}$, i.e.,
\begin{equation} \label{eq_lmb_fusion}
    \Tilde{\bm{\pi}}(\bm{X}) = \dfrac{\prod_{i=1}^N \big(\bm{\pi}_i(\bm{X})\big)^{\omega_i} }{\underset{\bm{X}\in\mathbb{X}\times\mathbb{L}}{\sup} \prod_{i=1}^N \big(\bm{\pi}_i(\bm{X})\big)^{\omega_i}},
\end{equation}
where $\Tilde{\bm{\pi}} = \Big\{ \big( \Tilde{\tau}^{(\ell)}, \Tilde{\gamma}^{(\ell)}, \Tilde{f}^{(\ell)} \big) \Big\}_{\ell\in\mathbb{L}}$, and
\begin{align}
    \Tilde{\tau}^{(\ell)} &= \frac{ \prod_{i=1}^N \big( \tau_i^{(\ell)} \big)^{\omega_i} }{\max\Big\{ \prod_{i=1}^N \big( \tau_i^{(\ell)} \big)^{\omega_i}, ~ \prod_{i=1}^N \big( \gamma_i^{(\ell)} \big)^{\omega_i} \eta_f \Big\}}, \\
    \Tilde{\gamma}^{(\ell)} &= \frac{  \prod_{i=1}^N \big(\gamma_i^{(\ell)} \big)^{\omega_i} \eta_f }{\max\Big\{ \prod_{i=1}^N \big( \tau_i^{(\ell)} \big)^{\omega_i} , ~ \prod_{i=1}^N \big( \gamma_i^{(\ell)} \big)^{\omega_i} \eta_f \Big\}}, \\
    \Tilde{f}^{(\ell)}(x) &= \frac{1}{ \eta_f } \prod_{i=1}^N \Big(f_i^{(\ell)}(x) \Big)^{\omega_i}, \\
    \eta_f & = \sup_{x'\in\mathbb{X}} \prod_{i=1}^N \Big(f_i^{(\ell)}(x') \Big)^{\omega_i},
\end{align}
where the fusion weight $\omega_i \in[0, 1]$ and $\max_{1\leq i \leq N} \omega_i = 1$. It is possible that each fusion weight $\omega_i=1$ in the case when all the information of LMB UFSs is independent. 
\end{prop}

Suppose two labeled Bernoulli UFSs $\{(\tau_1^{(\ell)}, \gamma_1^{(\ell)}, f_1^{(\ell)}) \}_{\ell \in \mathbb{L}}$ and $\{(\tau_2^{(\ell)}, \gamma_2^{(\ell)}, f_2^{(\ell)}) \}_{\ell \in \mathbb{L}}$ from sensor 1 and sensor 2 that represent the same target $\ell$. The possibility functions $f_1^{(\ell)}$ and $f_2^{(\ell)}$ of two LMB UFSs are modeled as GMMs 
\begin{align}
    f_1^{(\ell)}(x) &= \underset{1\leq i\leq N_1}{\max} w_1^{(i)} \bar{\mathcal{N}}\big(x; \mu_1^{(i)}, \Sigma_1^{(i)}\big) \\
    f_2^{(\ell)}(x) &= \underset{1\leq j\leq N_2}{\max} w_2^{(j)} \bar{\mathcal{N}}\big(x; \mu_2^{(j)}, \Sigma_2^{(j)}\big). 
\end{align}
The fused possibility function $\Tilde{f}^{(\ell)}$ of $(f_1^{(\ell)})^{\omega_1}$ and $(f_2^{(\ell)})^{\omega_2}$ is still in the form of GMM, and this can be derived based on the property that the exponentiation of a GMM is still a GMM~\footnote{In the GCI method, the exponentiation of a Gaussian mixture does not remain the form of Gaussian mixture, and an approximation is required to preserve the GM form. This assumption is only reasonable if the Gaussian components are well separated. In contrast, the exponentiation of a GMM remains the GMM forms without using any approximation. }:
\begin{equation}
    \Big[\max_{1\leq i\leq N} w_i \bar{\mathcal{N}}\big(x; \mu_i, \Sigma_i\big) \Big]^{\omega} = \max_{1\leq i\leq N} w_i^{\omega} \bar{\mathcal{N}}\big(x; \mu_i, \frac{\Sigma_i}{\omega}\big).
\end{equation}
In addition, the product of two weighted Gaussian possibility functions is a weighted Gaussian possibility function, i.e., 
\begin{equation}
    w_1 \bar{\mathcal{N}}\big(x; \mu_1, \Sigma_1\big) \cdot w_2 \bar{\mathcal{N}}\big(x; \mu_2, \Sigma_2\big) = w_{12} \bar{\mathcal{N}}\big(x; \mu_{12}, \Sigma_{12}\big), 
\end{equation}
where
\begin{equation}
    \begin{split}
        w_{12} & = w_1 w_2 \bar{\mathcal{N}}\big(\mu_1; \mu_2, \Sigma_1+\Sigma_2\big) \\
        \Sigma_{12} & = \big(\Sigma_1^{-1} + \Sigma_2^{-1} \big)^{-1} \\
        \mu_{12} & = \Sigma_{12}\big(\Sigma_1^{-1} \mu_1 + \Sigma_2^{-1} \mu_2\big).
    \end{split}
\end{equation}
Therefore, the product of the two GMMs $f_1^{(\ell)}$ and $f_2^{(\ell)}$ can be obtained as the pair-wise product of all weighted max-mixture components, which is also a GMM with $N_1 N_2$ components.

Based on the above properties and $\sup_{x \in S} \bar{\mathcal{N}}(x; \mu, \Sigma) = 1$, $\Tilde{f}^{(\ell)}(x)$ can be rewritten as the following form:
\begin{equation} \label{eq_GMM_EMF}
    \Tilde{f}^{(\ell)}(x) = \frac{1}{\eta_f } \underset{1\leq i\leq N_1}{\max} \underset{1\leq j\leq N_2}{\max} w_{12}^{(ij)} \bar{\mathcal{N}}\big(x; \mu_{12}^{(ij)}, \Sigma_{12}^{(ij)}\big), \\ 
\end{equation} 
where 
\begin{equation}
    \begin{split}
        \eta_f &= \underset{1\leq i\leq N_1}{\max} \underset{1\leq j\leq N_2}{\max} w_{12}^{(ij)}\\
        w_{12}^{(ij)} &= (w_1^{(i)})^{\omega_1} (w_2^{(j)})^{\omega_2} \bar{\mathcal{N}}\Big( \mu_1^{(i)}-\mu_2^{(j)};0, \frac{\Sigma_1^{(i)}}{\omega_1} + \frac{\Sigma_2^{(j)}}{\omega_2} \Big) \\
        \Sigma_{12}^{(ij)} &= \big[\omega_1(\Sigma_1^{(i)})^{-1} + \omega_2(\Sigma_2^{(j)})^{-1} \big]^{-1}\\
        \mu_{12}^{(ij)} &= \Sigma_{12}^{(ij)} \big[\omega_1(\Sigma_1^{(i)})^{-1}\mu_1^{(i)} + \omega_2(\Sigma_2^{(j)})^{-1}\mu_2^{(j)} \big]. 
    \end{split}
\end{equation}
The fusion formula may yield a large number of new max-mixtures. The pruning and merging scheme is effective for reducing the number of fused max-mixtures. Alternative efficient implementations of the fusion method, e.g., partial consensus fusion method~\cite{li2017distributed}, can be applied to only exchange and fuse the significant components so as to avoid producing a large number of trivial max-mixtures. In addition, the fusion formula can be further applied to $N$ sensors by sequentially applying the pairwise fusion $N-1$ times.

\subsection{LMB Fusion with Label Matching} \label{Sec_label}

Note that Proposition~\ref{prop_fusion} requires the label space to be synchronized across the sensor network. However, each sensor may have a distinct label space in applications, which prevents us from using Eq.~\eqref{eq_lmb_fusion} to perform LMB fusion. Therefore, the label-matching problem needs to be considered to determine whether Bernoulli UFSs from different sensors are referring to the same targets before performing the LMB fusion.

Several solutions have been proposed recently in response to the challenging label-matching problem. In this paper, we aim to tackle the label-matching issue from a slightly distinct perspective compared to the existing literature. Considering the FOVs of local sensors are not identical and the information of newborn targets is measurement-driven, the labels assigned to newborn targets by different sensors may not coincide. To address this issue, the proposed fusion methodology encompasses three sequential steps.

\textbf{1) Multi-Bernoulli Transformation.} Suppose the LMB UFSs are fused pairwise in either centralized or distributed fusion. Taking the fusion of two LMB UFSs $\bm{\pi}_{1,b} =\{(\tau_{1,b}^{(\ell)}, \gamma_{1,b}^{(\ell)}, f_{1,b}^{(\ell)}) \}_{\ell \in \mathbb{B}_1}$ and $\bm{\pi}_{2,b}=\{(\tau_{2,b}^{(\ell)}, \gamma_{2,b}^{(\ell)}, f_{2,b}^{(\ell)}) \}_{ \ell \in \mathbb{B}_2}$ regarding the newborn targets of sensor $1$ and sensor $2$ as an example, these two LMB UFSs are transformed to multi-Bernoulli UFSs, i.e., $\pi_{1,b}=\{(\tau_{1,b}^{(i)}, \gamma_{1,b}^{(i)}, f_{1,b}^{(i)}) \}_{i=1}^m$ and $\pi_{2,b}=\{(\tau_{2,b}^{(j)}, \gamma_{2,b}^{(j)}, f_{2,b}^{(j)}) \}_{j =1}^n$, respectively. 

\textbf{2) Multi-Bernoulli UFS fusion.}  The association of newborn targets from $\pi_{1,b}$ and $\pi_{2,b}$ is computed by assessing their distance based on $\eta_f$, where $\eta_f(i,j)$ represents the similarity of the Bernoulli UFS $i$ and $j$ from sensor 1 and sensor 2, respectively. A threshold $T_n$ is defined to determine the association of two Bernoulli UFSs. However, it is possible that a Bernoulli UFS in $\pi_{1,b}$ is close to multiple Bernoulli UFSs in $\pi_{2,b}$, meaning that the association of Bernoulli UFSs cannot be solely determined by using a user-defined threshold $T_n$. 

\begin{figure}[h!] 
\centering
\includegraphics[width=\linewidth]{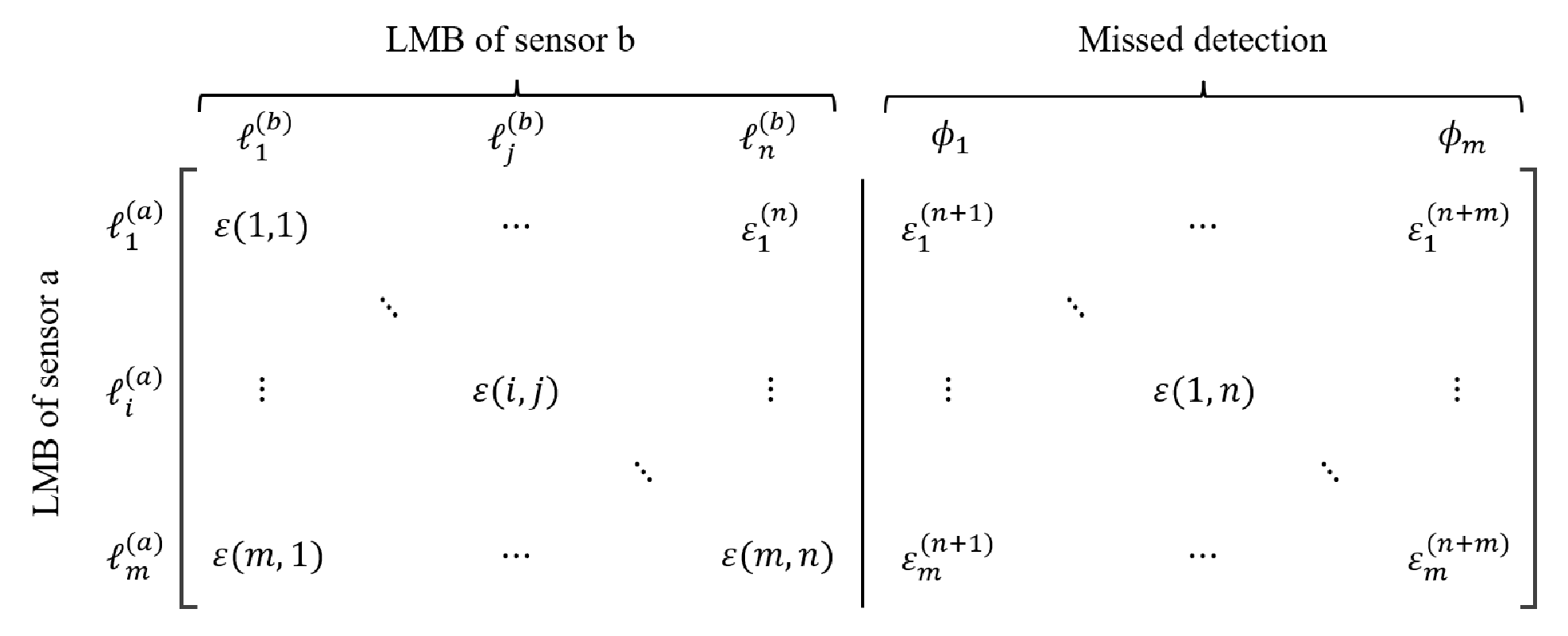}
\caption{The cost matrix of association} \label{Fig_cost}
\end{figure}

Alternatively, the association of $\pi_{1,b}$ and $\pi_{2,b}$ is formulated as a rank assignment problem in this paper. Enumerating $\pi_{1,b}=\{(\tau_{1,b}^{(i)}, \gamma_{1,b}^{(i)}, f_{1,b}^{(i)}) \}_{i=1}^m$ and $\pi_{2,b}=\{(\tau_{2,b}^{(j)}, \gamma_{2,b}^{(j)}, f_{2,b}^{(j)}) \}_{j =1}^n$ yields an $m\times n$ cost matrix $C$, where each entry can be computed based on $\eta_f(i,j)$. Moreover, the cost matrix $C$ is augmented by taking into account the missed detection $\phi$ of all Bernoulli UFSs in $\pi_{1,b}$. The missed detection of a newborn target is modeled by a Bernoulli UFS $X_{\phi}$ with parameter set $(\tau_{\phi}, \gamma_{\phi}, f_{\phi})$, where $\tau_{\phi}= r_0$ is a constant represents the possibility of non-existence, $\gamma_{\phi}({x})= r_1 d_f({x})$ consisting a constant $r_1$ denoting the possibility of existence and a state-dependent $d_f({x})$ represents the possibility of detection failure, and $f_{\phi}(\cdot)$ is a uniform possibility function that models our ignorance about the spatial distribution of undetected targets. As shown in Fig.~\ref{Fig_cost}, the cost matrix is an $m\times (n+m)$ optimal assignment matrix, where the cost of assigning the Bernoulli UFS $X_i^{(1)}$ to $X_j^{(2)}$ is given by $C(i,j)$, i.e., 
\begin{equation}
    C(i,j)= \begin{cases*}
      -\ln \Big( \gamma_{1,b}^{(i)} \gamma_{2,b}^{(j)} \eta_f(i,j) \Big) \quad & if $j \leq n$ \\
      -\ln(\gamma_{1,b}^{(i)} \gamma_{\phi} \eta_f(i,j)) & if $j> n$.
    \end{cases*} 
\end{equation}
Moreover, the threshold $T_n$ can still be applied to eliminate trivial associations, i.e., $\eta_f(i,j) = 0$ if $\eta_f(i,j)< T_n$, and therefore simplifies the complexity of the optimization. 

This optimal assignment problem can be efficiently addressed by several optimization algorithms, and Murty's method is employed in this paper. The solution is a matrix $S$ that represents the association of Bernoulli UFSs. On top of that, the associated Bernoulli UFSs can then be fused using Eq.~\eqref{eq_lmb_fusion}, and the fusion of all associated LMB UFSs generates a multi-Bernoulli UFS denoted by $\Tilde{\pi}_{12,b}$. The fusion of all Bernoulli UFSs in $\pi_{1,b}$ with missed detection yields a multi-Bernoulli UFS $\Tilde{\pi}_{1\phi,b}$. Moreover, if a Bernoulli UFS in $\hat{\bm{\pi}}_{2,b}$ is not associated with any components in $\hat{\bm{\pi}}_{1,b}$, it should associate with a missed detection. This type of association yields a multi-Bernoulli UFS denoted by $\Tilde{\pi}_{\phi 2,b}$. Fusing $\pi_{1,b}$ and $\pi_{2,b}$ generates a multi-Bernoulli UFS, which is given by 
\begin{equation}
    \Tilde{\pi}_b = \Tilde{\pi}_{12,b} \cup \Tilde{\pi}_{1\phi,b} \cup \Tilde{\pi}_{\phi2,b}.
\end{equation}

\textbf{3) LMB UFS Recovery.} The fused multi-Bernoulli UFS $\Tilde{\pi}_b$ is then transformed to an LMB UFS $\Tilde{\bm{\pi}}_b$ by assigning distinct labels to each Bernoulli UFS in $\Tilde{\pi}_b$ at the current sensor node.   

The above process can also be applied to handle the fusion of both the newborn target and the existing target in distributed fusion. For the fusion of existing targets in the centralized network, the association process is avoided since the label of existing targets is synchronized across the sensor network. Note that missed detection still needs to be considered in fusing existing targets since the components of LMB UFSs from different sensors may not be identical.

%=================================================================
\section{Possibility Multi-Sensor Multi-Target Tracking} \label{Sec_implementation}

In this section, the proposed possibility LMB filter and the fusion principle of LMB UFS are combined to investigate MSMTT algorithms in centralized and distributed sensor networks. The topology of the two types of sensor networks is introduced in Section~\ref{Sec_network}, followed by the discussion of the implementation aspects of centralized and distributed MSMTT presented in Section~\ref{Sec_centralized} and Section~\ref{Sec_centralized}, respectively.

\subsection{Sensor Network} \label{Sec_network}

To depict the communication structure of a sensor network, a graph $\mathcal{G}={\mathcal{V}, \mathcal{E} }$ can be employed, where $\mathcal{V}={1, \cdots, N }$ denotes the set comprising all sensor nodes, and $\mathcal{E} \subset \mathcal{V} \times \mathcal{V}$ denotes the edge set that characterizes the communication links between these nodes. Specifically, an edge $\mathcal{E}(i,j)$ signifies that node $j$ is capable of receiving information from node $i$. For a sensor node $i\in \mathcal{V}$, $\mathcal{V}_i$ denotes the set of neighboring nodes that can send information to $i$, and $\mathcal{E}(i,i) \in \mathcal{V}_i$.

\begin{figure}[h!]
    \centering
    \includegraphics[width=\linewidth]{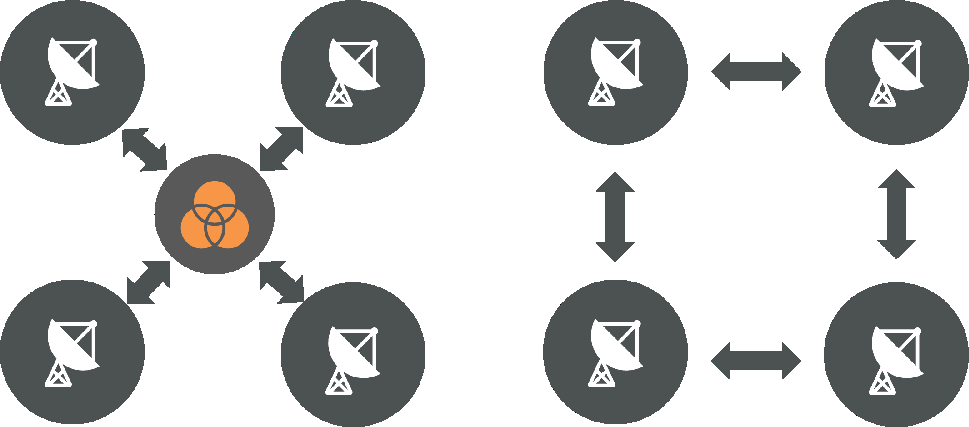}
    \caption{Sensor network} \label{Fig_network}
\end{figure}%

The left subfigure of Fig.~\ref{Fig_network} serves as an illustrative instance of a centralized sensor network topology comprising four sensors. In this topology, the network adopts an undirected graph representation, with bidirectional communication links shown as double-headed arrows. Notably, a central node assumes the responsibility of disseminating and processing information across the entire network, enabling seamless communication with all interconnected nodes. In contrast, the right subfigure of Fig.~\ref{Fig_network} depicts a distributed sensor network topology with the same number of sensors. Similarly represented as an undirected graph, this topology lacks a central node capable of collecting information from each local node within the network. 

The centralized network benefits from its ability to access and analyze data from across the network comprehensively. On the other hand, the distributed network offers advantages in terms of enhanced flexibility and robustness. The selection of an appropriate network topology necessitates careful consideration of specific application requirements. Therefore, it is imperative to investigate MSMTT methods within the OPM framework for network topologies, to provide valuable insights into the performance characteristics and suitability of these two network configurations for various applications.

\subsection{Centralized MSMTT} \label{Sec_centralized}

The centralized MSMTT can be accomplished through a three-step process, which includes centralized prediction, localized update, and centralized fusion.  

In the centralized prediction, the LMB prediction is performed at the central node to predict multi-target densities to the current time step. Assuming the multi-target dynamical model is linear and the distribution of process noise variations is Gaussian, the transition function $f(x_+|x,\ell)= \bar{\mathcal{N}}(x_+; Fx, Q)$ is Gaussian.  In this expression, $F$ represents the state transition matrix and $Q$ denotes the variance of the process noise. Suppose the possibility function $f(x, \ell)$ including the fused birth intensity measure is a GMM: 
\begin{equation}
    f(x, \ell) = \underset{1\leq i\leq N{(\ell)}}{\max} w_i{(\ell)} \bar{\mathcal{N}} \big(x; \mu_i{(\ell)}, \Sigma_i{(\ell)}\big),
\end{equation}
the predicted possibility function $f_+(x, \ell)$ can also be represented as a GMM
\begin{align}
    f_s(x, \ell) & = \underset{1 \leq i\leq N{(\ell)}}{\max} w_i{(\ell)} \bar{\mathcal{N}} \big(x; \mu_{s,i}{(\ell)}, \Sigma_{s,i}{(\ell)} \big)\\
    \mu_{s,i}{(\ell)} & = F \mu_i{(\ell)},\\
    \Sigma_{s,i}{(\ell)} & = F \Sigma_i{(\ell)} F^T +Q.
\end{align}
Assuming that $\lambda_s(x, \ell)$ and $\lambda_d(x, \ell)$ are independent of both the state $x$ and the label $\ell$, i.e., $\lambda_s(\ell) =\lambda_s(x, \ell)$ and $\lambda_d(\ell) = \lambda_d(x, \ell)$. Then, the target survival possibility and non-survival possibility can be rewritten as $\eta_s^{(\ell)} = \lambda_s(\ell)$ and $\eta_d^{(\ell)} = \lambda_d(\ell)$, respectively.

In the local update, the key to enabling the full exploitation of the observations of all sensors lies in maintaining the independence between sensor nodes. To guarantee the local information is independent of each other, the LMB update should be carried out solely at each node to process local information. Given a Gaussian likelihood function $l(z|x,\ell) = \bar{{\mathcal{N}}} (z; Hx, R)$, with $H$ denoting the observation matrix and $R$ representing the observation noise variance, it is assumed that the possibilities of successful detection $d_s(x,\ell)$ and detection failure $d_f(x,\ell)$ are independent of both the state $x$ and the label $\ell$. Furthermore, considering a single target distribution represented by a GMM,
\begin{equation}
    f_+(x,\ell) =  \underset{1 \leq i\leq N(\ell)}{\max} w_i(\ell) \bar{\mathcal{N}} \big(x; \mu_i(\ell) , \Sigma_i(\ell) \big),
\end{equation}
the absence possibility, presence possibility, and the posterior possibility function of $\ell$ given the measurement $Z_j$ of sensor $j$ are given by
\begin{align}
    \tau^{(\ell)} & = \max_{(I_+,\theta)\in\mathcal{F}(\mathbb{L}_+)\times\Theta_+} (1- 1_{I_+}(\ell)) w^{(I_+,\theta)}(Z_j) \label{eq_cen_tau}\\
    \gamma^{(\ell)} & = \max_{(I_+,\theta)\in\mathcal{F}(\mathbb{L}_+)\times\Theta_+} 1_{I_+}(\ell) w^{(I_+,\theta)}(Z_j) \\
    f(x, \ell|Z_j) &= \frac{1}{\gamma^{(\ell)}} \max_{(I_+,\theta)\in\mathcal{F}(\mathbb{L}_+)\times\Theta_+}  w^{(I_+, \theta)}(Z_j)  1_{I_+}(\ell)  \notag\\ 
    & \quad \times \max_{1 \leq i\leq N(\ell)} w^{(\theta)}_i(\ell)  \bar{\mathcal{N}} \Big(x; \mu^{(\theta)}_{i}(\ell), \Sigma^{(\theta)}_{i}(\ell) \Big),
\end{align}
where
\begin{align}
    w^{(I_+, \theta)}(Z_j) & = w^{(I_+)} \begin{cases}
      \frac{d_s \bar{\mathcal{N}} \big(z; \hat{z}^{(\theta)}_i, \Sigma^{(\theta)}_{z,i}(\ell) \big)}{\kappa(z_{\theta(\ell)})}, &  \text{if} ~ \theta(\ell) \neq 0  \\
      d_f & \text{if} ~ \theta(\ell)=0
    \end{cases} \\
    w^{(\theta)}_i(\ell) & \propto w_i(\ell) \bar{\mathcal{N}} \big(z; \hat{z}^{(\theta)}_i, \Sigma^{(\theta)}_{z,i}(\ell) \big) \\
    \mu^{(\theta)}_{i}(\ell) &= 
    \begin{cases}
      \mu^{(\theta)}_i+ K_i^{(\theta)}(\ell) (z- \hat{z}^{(\theta)}_i)  & \text{if} ~ \theta(\ell) \neq 0 \\
     \mu^{(\theta)}_i & \text{if} ~ \theta(\ell)=0
    \end{cases} \\
    \Sigma^{(\theta)}_i(\ell) &=
    \begin{cases}
        [I - K_i^{(\theta)}(\ell) H] \Sigma^{(\theta)}_i(\ell) & \text{if} ~ \theta(\ell) \neq 0\\
        \Sigma^{(\theta)}_i(\ell) & \text{if} ~ \theta(\ell)=0 
    \end{cases} \\
    \Sigma^{(\theta)}_{z,i}(\ell) & = H \Sigma^{(\theta)}_i(\ell) H^T +R \\
    K_i^{(\theta)}(\ell) & = \Sigma^{(\theta)}_i(\ell) H^T \big( \Sigma^{(\theta)}_{z,i}(\ell) \big)^{-1}. \label{eq_cen_sigma}
\end{align}
The pruning and merging scheme need to be performed such that insignificant Gaussian components can be removed and similar Gaussian are combined to reduce the computational complexity. 

Once the local LMB update is complete, all sensors send their posterior LMB to the fusion center for centralized fusion. Since all local information is independent, the centralized fusion formula can be rewritten to the following form based on Eq.~\eqref{eq_lmb_fusion}, i.e., 
\begin{equation} \label{eq_cen_fusion}
    \Tilde{\bm{\pi}}(\bm{X}) = \dfrac{\prod_{i \in \mathcal{V}} \bm{\pi}_i(\bm{X}) }{\underset{\bm{X}\in\mathbb{X}\times\mathbb{L}}{\sup} \prod_{i \in \mathcal{V}} \bm{\pi}_i(\bm{X}) }.
\end{equation}
Compared to Eq.~\eqref{eq_lmb_fusion}, the fusion weights are all equal to one. The fused state will be fed back to each node to replace the local posterior for the subsequent recursive estimation.

\subsection{Distributed MSMTT} \label{Sec_distributed}

In order to maximize the utilization of observations from all sensors, it is crucial to maintain independence among sensor nodes during distributed fusion. To ensure the independence of local information, two steps must be taken in distributed MSMTT. Firstly, the Markov transition process in prediction needs to be discounted. Secondly, measurement updates should be performed independently at each node.

In the prediction, the discount strategy of the Markov transition process $\bm{\phi}(\bm{X}_+ | \bm{X})$ is performed by taking the power $\omega$ of each term~\cite{Houssineau2023}. Specifically, the absence possibility, presence possibility, and predicted covariance of the survival targets are given by 
\begin{align}
    \tau_s^{(\ell)} & = \max \{ \tau^{(\ell)}, \gamma^{(\ell)} (\eta_d(\ell))^{\omega}\},\\
    \gamma_s^{(\ell)} & = \gamma^{(\ell)} (\eta_s(\ell))^{\omega},\\
    \Sigma_{s,i}{(\ell)} & = F \Sigma_i{(\ell)} F^T +\frac{Q}{\omega}.
\end{align}
The weight and mean of the predicted GMM remain the same as the centralized case. 

In the measurement update, the LMB filter is running at each node to generate a local posterior of multi-target density. This process is identical to that of the centralized MSMTT through Eqs.~\eqref{eq_cen_tau}-~\eqref{eq_cen_sigma}.

In the distributed consensus fusion, a sensor $i$ can exchange information with its neighboring nodes $j \in \mathcal{V}_i$ and achieve the LMB fusion using the following equation. 
\begin{equation} \label{eq_dis_fusion}
    \Tilde{\bm{\pi}}(\bm{X}) = \dfrac{\prod_{j \in \mathcal{V}_i} \big( \bm{\pi}_j(\bm{X}) \big)^{\omega_{ij}} }{\underset{\bm{X}\in\mathbb{X}\times\mathbb{L}}{\sup} \prod_{j \in \mathcal{V}_i} \big( \bm{\pi}_i(\bm{X}) \big)^{\omega_{ij}} }.
\end{equation}
A necessary condition is that the weights of all sensors in $\mathcal{V}_i$ are normalized to one, i.e., $\sum_{j \in \mathcal{V}_i} \omega_{i,j}=1$. All fusion weights form a matrix $W$, where the $(i,j)$th entry represents $\omega_{ij}$. The popular Metropolis weight is employed in this study, which forms the following matrix~\cite{xiao2005scheme}:
\begin{equation}
  \omega_{i,j} =
    \begin{cases}
      \frac{1}{1+ \max\{ N_i,N_j \}} & \text{if $(i,j) \in \mathcal{E} $}\\
      1- \sum_{(i,j) \in \varepsilon} \omega_{i,j}  & \text{if $i=j$}\\
      0 & \text{otherwise},
    \end{cases}       
\end{equation}
where $N_i= |\mathcal{V}_i|$ represents the degree or cardinality of sensor $i$, i.e, the number of neighboring nodes of sensor $i$. Extensive research has demonstrated the achievability of a global consensus within a finite number of iterations.~\cite{farahmand2011set,li2017distributed}.

%=================================================================
\section{Simulation} \label{sec_Test}

\subsection{Simulation Design}

To validate the developed possibilistic centralized and distributed multi-sensor fusion methods, two simulated MSMTT scenarios are proposed in this section. Moreover, typical centralized and distributed multi-sensor fusion methods using the probabilistic LMB filter~\cite{reuter2014labeled} are also tested for the sake of comparison. 

The mission of the two MSMTT scenarios is to jointly track multiple targets within a 2-dimensional Monitoring area $S= [-1000, 1000]~\text{m} \times [-1000, 1000]~\text{m}$. The state $x=[x_1, \dot{x}_1,  x_2, \dot{x}_2]$ of a target is modeled as an uncertain variable. The entire tracking time period is 100 evenly discretized time steps with a time interval of $\Delta_t = 1$ s. The state of targets evolves based on a linear dynamical model, which is modeled as a Gaussian PDF $f_{k|k-1}(x|x_{k-1}) = \bar{\mathcal{N}}{\big(x; F_k x_{k-1}, Q_k \big)}$, where the state transition matrix $F_k$ and the process noise matrix $Q_k$ are given by  
\begin{gather*} \label{eq_F}
F_k = 
  \begin{bmatrix}
   I_2 & I_2 \otimes \Delta_t \\
   0 & I_2
   \end{bmatrix}, ~ 
Q_k = \frac{\sigma_q^2 I_2 }{4} \otimes
   \begin{bmatrix}
   \Delta_t^4 & 2 \Delta_t^3 \\
   2 \Delta_t^3 & \Delta_t^4
   \end{bmatrix},
\end{gather*}
where $\sigma_q$ is the Standard Deviation (STD) of the process noise, $I_2=[1,1]^T$, and $\otimes$ represents the Kronecker product. 

Given the measurement $z_k$ and a single-target state $x_k$ at time $k$, the likelihood function is modeled as a Gaussian PDF $l_{k}(z_k|x_k) = \bar{\mathcal{N}}{\big(z_k; H(x_k-x_{s,k}), R_k \big)}$, where $x_{s,k}$ is the state of sensor $s$ at time $k$. Suppose the sensors in simulation can detect the position of targets, and the measurement noise is defined as $R_k= \sigma_r^2 I_2$ and $\sigma_r=5$ m is the STD of measurement noise. The probability of detection is a Gaussian possibility function $p_{d,k}(x) = \bar{\mathcal{N}}{(Hx; Hx_{s,k}, \sigma_s^2 I_2)}$, where the STD $\sigma_s$ denotes the detection capability of sensors. Clutters are generated based on $\bar{\mathcal{N}}(\cdot; Hx_{s,k}, \sigma_s^2I_2)$. The occurrence of false alarms is simulated based on a Poisson distribution with an expected value $\lambda_{fa}= 10$. All tested targets are generated based on the measurement-driven birth model.

For the possibilistic multi-sensor fusion method, suppose the information regarding the dynamical model is not perfectly known, the state transition function is represented by a Gaussian possibility function, i.e., $\bar{\mathcal{N}}{\big(x_k; F_k x_{k-1}, Q_k \big)}$. The possibilities of target survival and death are set to $\lambda_s = 1$ and $\lambda_d = 0.05$, respectively. Likewise, the likelihood function is defined as $\bar{\mathcal{N}}{\big(z_k; H(x_k-x_{s,k}), R_k \big)}$. Assume the possibility of successful detection equals one $d_{s,k}(x)=1$, and the detection failure is given by $d_{f,k}(x)=1 - \bar{\mathcal{N}}{(x; x_{s,k}, \sigma_s^2 I_2)}$. The possibility of false alarms is calculated as 
\begin{equation}
    \kappa(z) = \frac{1}{V} 2\pi \sigma_r^2 \lambda_{fa} \bar{\mathcal{N}}(z, H x_{s,k}, \sigma_s^2I_2).
\end{equation}
The multi-target state is extracted from the posterior LMB using the MAP cardinality estimator. 

In the possibility LMB filter, the threshold of pruning is set to $1\times 10^{-3}$. The merging of Gaussian possibility functions is determined using the possibilistic Hellinger distance~\cite{houssineau2021linear}, where the merging threshold is set to $0.1$. Following the definition in Ref.~\cite{vo2014labeled}, the pruning threshold used in the typical LMB filter is defined as $1\times 10^{-4}$, and the Mahalanobis distance is utilized for merging with a threshold 4. All the tested methods are assessed through 100 Monte Carlo (MC) runs using different measurements. The multi-target tracking errors are evaluated by the Optimal Sub-Pattern Assignment (OSPA) distance metric~\cite{schuhmacher2008consistent}, where the cut-off parameter is $c=100~\text{m}$ and the order is $p=2$. 

% \begin{table}[h!]
% \begin{center}
% \caption{Filtering design parameters}
% \label{tab1}
% \begin{tabular}{| c | c | c |}
% \hline
% Parameters & Possibility LMB & Probability LMB \\
% \hline
% Pruning threshold & $1\times 10^{-4}$, 
% & $1\times 10^{-5}$\\
% \hline
% Merging threshold & 0.2 & 4 \\ 
% \hline
%  &  &  \\
% \hline 
% \end{tabular}
% \end{center}
% \end{table}

\subsection{Case A}

This section employs a network of 4 sensors fixed at the following location $x_{s,1}=[-1000, -1000]$ m, $x_{s,2}=[-1000, 1000]$ m, $x_{s,3}=[1000, -1000]$ m, and $x_{s,4}=[1000, 1000]$ m. The targets are born from 3 fixed birth locations, i.e., $[ 800; 500]$ m, $[ -800; 500]$ m, $[ 0; -800]$ m, from different times, and a target disappeared at time $50$. 

This case assumes the sensors' detection capability is limited, where the detection probability and detection possibility are computed based on the specification of $\sigma_s= 1000$ m. Furthermore, the same information about the new targets born is utilized in both possibilistic and probabilistic methods. The existence possibility of newborn targets is $\gamma_{b}= 1\times 10^{-3}$, and the non-existence possibility is $\tau_{b}=1$. The probability of existence is defined as $r_b= 1\times 10^{-3}$. 

% For the distributed sensor architecture, the Metropolis weight matrix of the 4 sensors is given by 
% \begin{equation}
%     W= \begin{bmatrix}
%     \frac{1}{3} & \frac{1}{3} & 0 & \frac{1}{3} \\[6pt]
%     \frac{1}{3} & \frac{1}{3} & \frac{1}{3} & 0 \\[6pt]
%     0 & \frac{1}{3} & \frac{1}{3} & \frac{1}{3} \\[6pt]
%     \frac{1}{3} & 0 & \frac{1}{3} & \frac{1}{3} \\
%     \end{bmatrix}.
% \end{equation}

\begin{figure}[h!] 
\centering
\includegraphics[width=\linewidth]{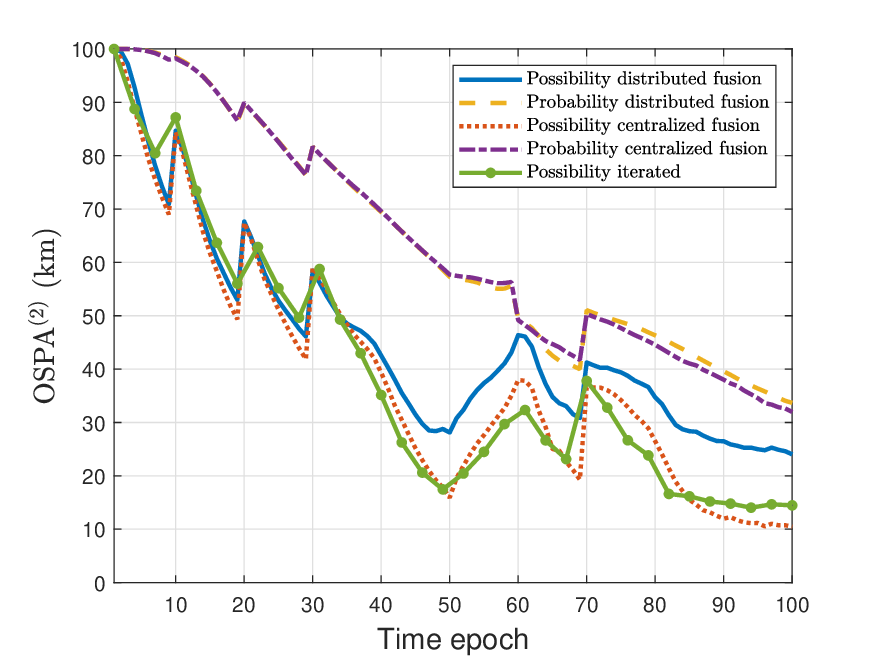}
\caption{OSPA$^{(2)}$ position distance of case A} \label{Fig_ospa1}
\end{figure}

The OSPA$^{(2)}$ position distances of the 4 tested multi-sensor fusion methods are shown in Fig.~\ref{Fig_ospa1}. Results show that the possibilistic centralized fusion yields the most accurate state estimation throughout the entire tracking process. In addition, the possibilistic distributed fusion is superior to the probabilistic fusion methods, and its result is comparable to the possibilistic centralized fusion. This disparity between the possibilistic and probabilistic fusion methods substantiates the enhanced robustness provided by the proposed approach in scenarios characterized by limited knowledge. To further validate the developed method, the iterated multi-sensor fusion method is further assessed, in which the measurements obtained from each sensor are sequentially processed using the possibilistic GLMB update step. The corresponding results, depicted as the green solid curve, reaffirm the ability of the possibilistic centralized fusion approach to achieve performance similar to that of the iterated fusion technique.

\begin{figure}[h!] 
\centering
\includegraphics[width=\linewidth]{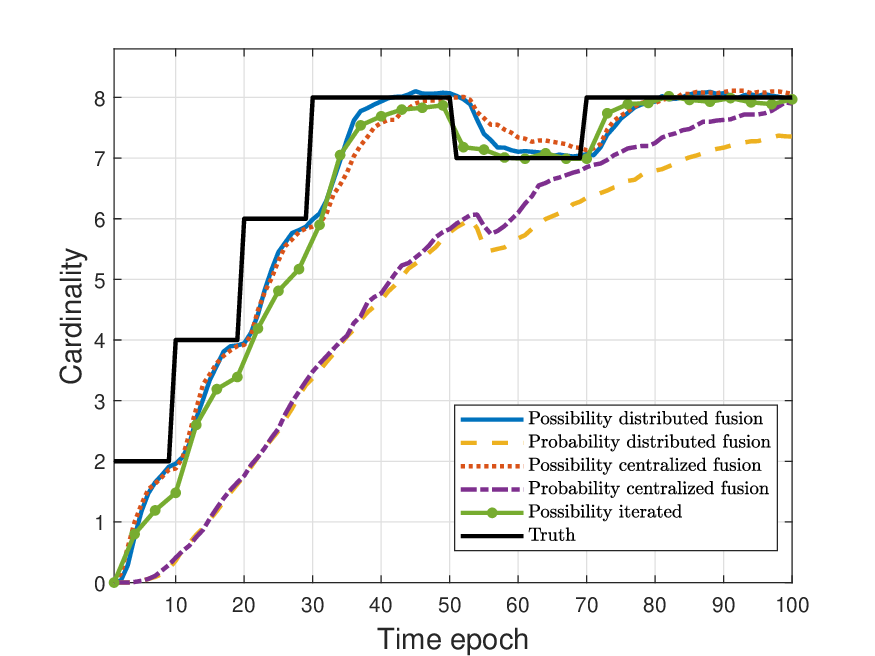}
\caption{Cardinality estimation of case A} \label{Fig_card1}
\end{figure}

The cardinality estimations of the 4 tested multi-sensor fusion methods are shown in Fig.~\ref{Fig_card1}. The obtained results exhibit a congruent trend with the corresponding OSPA distance outcomes. Notably, the two possibilistic fusion methods demonstrate a similar pattern in terms of cardinality estimation. Conversely, the probabilistic fusion methods display a relatively slower response to target births and deaths, particularly when the detection capability of sensors is limited. It is worth mentioning that, as the increase of the standard deviation $\sigma_s$, e.g., 2000 m, all fusion methods exhibit a tendency to yield comparable performance.

\subsection{Case B}

In this case study, the task is to track multiple targets in $S$ using 4 moving sensors. The FOV of each sensor is restricted to $\sigma_s= 500$ m. The initial positions of sensors are randomly generated within the area $S$, and their velocities have a consistent magnitude of $50$ m/s. If sensors reach the boundary of $S$, they will bounce back by turning an angle of $\pm 90^{\circ}$, guaranteeing that sensors move inside the surveillance region. 

The quantity of newborn targets is determined by a Poisson distribution characterized by an expected value of $\lambda_b = 0.3$. Additionally, no target births occur beyond the time instant $k=30$. The position $[x_1, x_2]$ of each target is randomly generated based on a uniform distribution in the monitoring area, and the velocity $[\dot{x}_1, \dot{x}_2]$ is generated randomly based on a Gaussian distribution with a zero mean and an STD $\sigma_v= 3$ m/s. In addition, the probability of newborn event is defined as $r_{B}= \lambda_{b} (2\pi \sigma_r^2) /V$, where $2\pi \sigma_r^2$ denotes the volume of the sensors' observation uncertainty, and $V= 2\pi \sigma_s^2$ indicates the volume of the whole observation space. Assuming both the possibilistic and probabilistic methods make use of the same information regarding the birth process.

\begin{figure}[h!] 
\centering
\includegraphics[width=\linewidth]{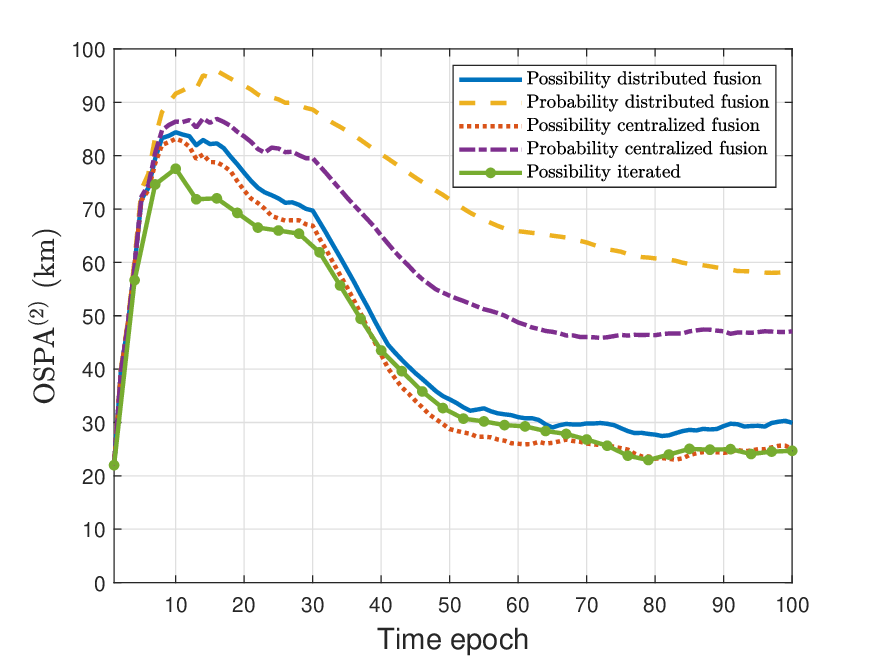}
\caption{OSPA$^{(2)}$ position distance of case B} \label{Fig_ospa2}
\end{figure}

\begin{figure}[h!] 
\centering
\includegraphics[width=\linewidth]{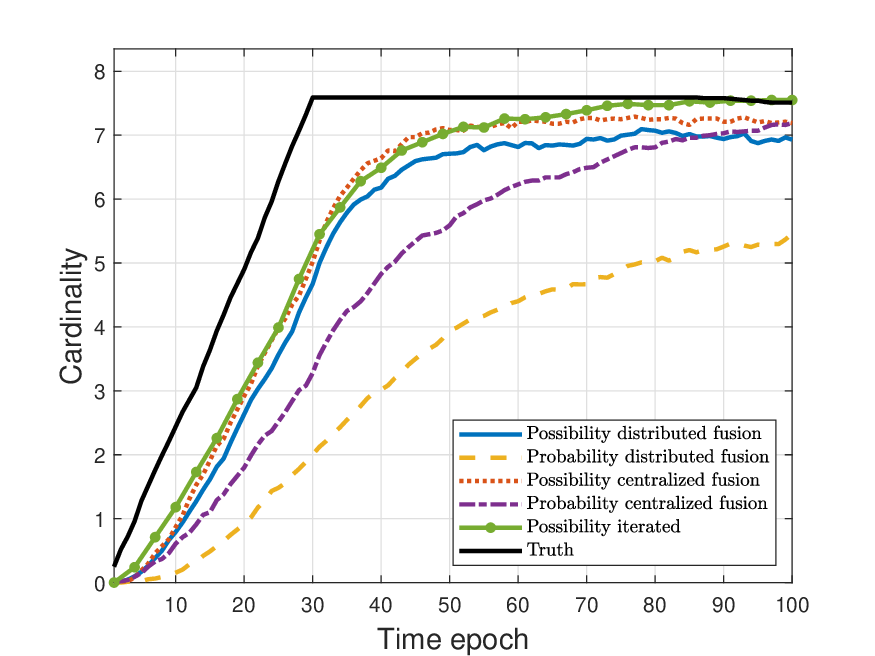}
\caption{Cardinality estimation of case B} \label{Fig_card2}
\end{figure}

Fig.~\ref{Fig_ospa2} depicts the OSPA$^{(2)}$ position distances of the four tested methods. Results validated that both possibilistic centralized and distributed fusion methods exhibit noteworthy superiority over their probabilistic counterparts. The performance of the possibilistic distributed fusion approach closely aligns with that of the possibilistic centralized fusion, while the probabilistic distributed fusion method encounters challenges in consistently maintaining the custody of certain targets. The observed disparities between the possibilistic and probabilistic outcomes underscore the enhanced robustness achieved by the proposed method, particularly when confronted with imperfect sensors and limited prior knowledge pertaining to target birth characteristics.

The cardinality estimations of the tested methods are shown in Fig.~\ref{Fig_card2}. Since the use of the birth time limit, the average number of targets existing is around 7.5 over 100 MC runs.  Notably, the two possibilistic fusion methods exhibit a rapid convergence toward the ground truth, and the distributed fusion approach yields comparable outcomes to that of centralized fusion. Conversely, the probabilistic fusion methods display significantly lower cardinality estimations compared to the ground truth, implying an inadequacy in typical MSMTT methods to accurately estimate the exact number of multiple targets. The superior performance exhibited by the proposed method can be attributed to its judicious modeling of key parameters such as the probability of detection and the probability of birth when confronted with limited information.

%=================================================================
\section{Conclusion}

This paper develops a possibilistic Labeled Multi-Bernoulli (LMB) Filter based on the novel concept of Outer Probability Measures (OPMs). The possibilistic LMB filter is further applied to formulate a multi-sensor fusion framework for the sake of improved robustness in Multi-Sensor Multi-Target Tracking (MSMTT). By utilizing OPMs, the proposed method ensures that the probabilistic parameters involved in the fusion process are appropriately constrained, thus accommodating limited information about their randomness. Furthermore, by assuming a shared label space across all sensors, the fusion formula of LMB UFS is derived to enable independent fusion of local information. This framework facilitates more effective utilization of local measurements and contributes to improved overall performance.

Two simulation scenarios are devised to evaluate the efficacy of the proposed method. The initial case study assumes fixed target births at predetermined locations, while the subsequent analysis examines a more formidable scenario involving four mobile sensors tracking newly generated targets within the surveillance region. The obtained simulation results confirm the superior accuracy of state and cardinality estimation achieved by the proposed method. This superiority is observed even in scenarios where there is imperfect information available regarding certain aspects of the MSMTT system.

It is important to acknowledge that the fusion of LMB UFSs in this study is predicated on the assumption of a congruent label space across the sensor network, which may not always be tenable in real-world applications. Consequently, future research endeavors will focus on investigating the fusion of LMB UFSs originating from sensors with disparate label spaces. This exploration aims to address the limitations associated with potential label space discrepancies and broaden the applicability of the proposed method.

\section*{Acknowledgments}
This work was funded by the National Natural Youth Fund (No.12202049), and the Beijing Institute of Technology Research Fund Program for Young Scholars.

\appendices

\section{} \label{App_LMB_predict}

\begin{proof} [\textbf{Proof of proposition \ref{prop_predict}}]

    Given the weight of the prior LMB UFS as the form of Eq.~\eqref{eq_w_LMB}, the weight $w_s(\cdot)$ has the following form   
    \begin{equation}
    \begin{split}
         w_s(L) &= \big[\eta_s\big]^L \max_{I \in \mathbb{L}} 1_I(L) \big[\eta_d \big]^{I-L} w(I) \\
         &= \big[\eta_s\big]^L \max_{I \supseteq L} \big[\eta_d \big]^{I-L} \prod_{i \in \mathbb{L}} \tau^{(i)} \prod_{\ell \in I} \frac{1_{\mathbb{L}}(\ell)\gamma^{(\ell)}}{ \tau^{(\ell)}} \\
         &= \big[\eta_s\big]^L \max_{I \supseteq L} \big[\eta_d \big]^{I-L} \big[\tau\big]^{\mathbb{L}} \Big[\frac{\gamma}{\tau}\Big]^{I} \\
         &= \big[\gamma \eta_s\big]^L \big[\tau\big]^{\mathbb{L}-L} \max_{I \supseteq L} \Big[\frac{\gamma \eta_d}{\tau}\Big]^{I-L}.
    \end{split}
    \end{equation}
    The maximum term in the last line of the above equation can be rewritten as 
    \begin{equation}
        \max_{I \supseteq L} \Big[\frac{\gamma \eta_d}{\tau}\Big]^{I-L} = \max_{J \subseteq \mathbb{L}-L} \Big[\frac{\gamma \eta_d}{\tau}\Big]^J = \Big[\frac{\gamma \eta_d}{\tau}\Big]^{J'}, 
    \end{equation}
    where $J'\subseteq \mathbb{L}-L$ indicating that the term $\Big[\frac{\gamma \eta_d}{\tau}\Big]^{J}$ reaches the maximum when $J= J'$. 
    
    Then, $w_s(L)$ can be rewritten as
    \begin{equation}  \label{eq_wS_proof}
        \begin{split}
            w_s(L) & = \big[\gamma \eta_s\big]^L \big[\tau\big]^{\mathbb{L}-L} \max_{I \supseteq L} \Big[\frac{\gamma \eta_d}{\tau}\Big]^{I-L} \\
            &= \big[\gamma \eta_s \big]^L \big[\tau\big]^{\mathbb{L}-L} \Big[\frac{\gamma \eta_d}{\tau}\Big]^{J'} \\
            &= \big[\gamma \eta_s \big]^L \big[\tau\big]^{\mathbb{L}-L-J'} \big[ \gamma \eta_d \big]^{J'}. \\
        \end{split}
    \end{equation}
    
    Note that for any $\ell \in J'$, the fraction should always equal to or greater than one, i.e., $\gamma^{(\ell)} \eta_d^{(\ell)} \geq \tau^{(\ell)}$, and for any $\ell \in \mathbb{L}-L-J'$, the fraction should always smaller than one, i.e., $\gamma^{(\ell)} \eta_d^{(\ell)} \leq \tau^{(\ell)}$. Therefore, we have 
    \begin{equation} \label{eq_qs_proof}
        \big[\tau\big]^{\mathbb{L}-L-J'} \big[ \gamma \eta_d \big]^{J'} = \big[\max\{ \tau, \gamma \eta_d\} \big]^{\mathbb{L}-L}
    \end{equation}
    Substituting Eq.~\eqref{eq_qs_proof} into Eq.~\eqref{eq_wS_proof},  the weight of survival objects can be rewritten as the standard LMB weight as Eq.~\eqref{eq_w_LMB}, i.e.,
\begin{equation}
    \begin{split}
         w_s(L) &=  \big[ \max\{ \tau, ~ \gamma \eta_d\} \big]^{\mathbb{L}-L} \big[\gamma \eta_s \big]^L \\
         & = \big[ \max\{ \tau, ~ \gamma \eta_d\} \big]^{\mathbb{L}} \Big[ \frac{\gamma \eta_s}{\max\{ \tau, ~ \gamma \eta_d\}} \Big]^L \\
         & = \big[ \tau_s\big]^{\mathbb{L}} \Big[ \frac{\gamma_s}{\tau_s} \Big]^L
    \end{split}
\end{equation}
In addition, as the label space of the posterior LMB and the birth LMB are disjoint $\mathbb{L}\cap\mathbb{B}=\phi$, the weight of the predicted LMB UFS is $w_+(I_+) = w_b(\mathbb{B} \cap I_+) w_s(\mathbb{L}\cap I_+ )$, and the predicted LMB UFS consisting multiple Bernoulli components as the combination of the survival LMB components and birth LMB components.

\end{proof}

\section{} \label{App_LMB_update}

\begin{proof}  [\textbf{Proof of proposition \ref{prop_update}}]

The LMB approximation of the multi-target posterior preserves the first-order moment of the $\delta$-GLMB UFS. The posterior $\delta$-GLMB is given by 
\begin{equation}
\begin{split}
    \bm{\pi}(Z|\bm{X}) = \Delta(\bm{X}) &\max_{(I_+,\theta) \in \mathcal{F}(\mathbb{L}_+)\times\Theta_+} w^{(I_+,\theta)}_Z \\
    & \times \delta_{I_+}(\mathcal{L}(\bm{X})) \Big[ f^{(\theta)}(\cdot|Z) \Big]^{\bm{X}}.
\end{split}
\end{equation}
The presence function of the $\delta$-GLMB is given by  
\begin{equation}
\begin{split}
    \bar{v}(x) & = \max_{(I_+,\theta)\in \mathcal{F}(\mathbb{L}_+)\times\Theta_+} \max_{\ell \in \mathbb{L}_+} f^{(\theta)}(x,\ell)  \\
    & \qquad \qquad \qquad \times \max_{L\subseteq \mathbb{L}_+} 1_L(\ell) w^{(I_+,\theta)}_Z \delta_{I_+}(L) \\
    & = \max_{\ell \in \mathbb{L}_+} \max_{(I_+,\theta)\in \mathcal{F}(\mathbb{L}_+)\times\Theta_+}  1_{I_+}(\ell) \\
    & \qquad \qquad \qquad \times w^{(I_+,\theta)}_Z f^{(\theta)}(x,\ell).
\end{split}
\end{equation}
% The inner maximum can be seen as the presence function of track $\ell$, i.e.,
% \begin{equation} 
%     \bar{v}^{(\ell)}(x)= \max_{(I,\theta)\in \mathcal{F}(\mathbb{L})\times\Theta}  1_I(\ell) w^{(I,\theta)} f^{(\theta)}(x,\ell). 
% \end{equation} 
% The set supremum of $\bar{v}^{(\ell)}(\cdot)$ results in the existence possibility of track $\ell$, i.e., $\gamma^{(\ell)} = \max_{(I,\theta)\in \mathcal{F}(\mathbb{L})\times\Theta}  1_I(\ell) w^{(I,\theta)}(Z)$, which is exactly the same as the posterior existence possibility. 

The presence function of an LMB UFS is given by 
\begin{equation} \label{eq_LMB_v}
    \bar{v}(x) = \max_{\ell \in \mathbb{L}_+} f(x,\ell) \max_{L\subseteq \mathbb{L}_+} 1_L(\ell) w(L). 
\end{equation}
The maximum over $L\subseteq \mathbb{L}_+$ can be rewritten as bellow
\begin{equation} \label{eq_max_w}
    \begin{split}
        \max_{L\subseteq \mathbb{L}_+} 1_L(\ell) w(L) &=\max_{L\subseteq \mathbb{L}_+} 1_L(\ell) \prod_{i\in\mathbb{L}_+} \tau^{(i)} \prod_{j\in L} 1_{\mathbb{L}_+}(j)\frac{ \gamma^{(j)}}{\tau^{(j)}} \\
        & = \max_{L\subseteq \mathbb{L}_+} 1_L(\ell) \prod_{i\in\mathbb{L}-L} \tau^{(i)} \prod_{j\in L}1_{\mathbb{L}}(j) \gamma^{(j)} \\
        & = \gamma^{(\ell)} \max_{L\subseteq \mathbb{L}_+} \Big[ 1_L(\ell)  \prod_{i\in\mathbb{L}-L} \tau^{(i)} \prod_{j\in L/\ell}\gamma^{(j)} \Big] \\
        & = \gamma^{(\ell)}. 
    \end{split}
\end{equation}
The indicator function $1_{\mathbb{L}}(j)$ is removed due to $j\in L\subseteq \mathbb{L}$, such that$1_{\mathbb{L}}(j)=1$. Due to the normalization $\max\{\tau, \gamma \}=1$, there always exists a hypothesis $L$ ensures that $\max_{L\subseteq \mathbb{L}} \Big[ 1_L(\ell)  \prod_{i\in\mathbb{L}-L} \tau^{(i)} \prod_{j\in L/\ell}\gamma^{(j)} \Big]=1$, such that the last line of the above equation holds true. 

Substituting Eq.~\eqref{eq_max_w} and the equations of $(r_{0,+}^{(\ell)}, r_{1,+}^{(\ell)}, f_+^{(\ell)})$ into Eq.~\eqref{eq_LMB_v} results in the detailed form of the presence function
\begin{equation}
\begin{split}
    \bar{v}(x) & = \max_{\ell \in \mathbb{L}_+} \max_{(I_+,\theta)\in\mathcal{F}(\mathbb{L}_+)\times\Theta_+}  \frac{1}{\gamma^{(\ell)}} \\
    & \qquad \qquad \times 1_{I}(\ell) w^{(I_+,\theta)}f^{(\theta)}(x,\ell) \gamma^{(\ell)}\\
    &= \max_{\ell \in \mathbb{L}_+} \max_{(I_+,\theta)\in\mathcal{F}(\mathbb{L}_+)\times\Theta_+}  1_{I_+}(\ell) w^{(I_+,\theta)}f^{(\theta)}(x,\ell).
\end{split}
\end{equation}
It is clear that the presence function of LMB is exactly identical to the presence function of $\delta$-GLMB given in Eq.~\eqref{eq_dGLMB_v}. 
\end{proof}

\section{} \label{App_fusion}

\begin{proof} [\textbf{Proof of proposition \ref{prop_fusion}}]
The product of two LMB UFSs $\bm{\pi}_1=\{(\tau_1^{(\ell)}, \gamma_1^{(\ell)}, f_1^{(\ell)}) \}_{\ell\in\mathbb{L}}$ and $\bm{\pi}_2=\{(\tau_2^{(\ell)}, \gamma_2^{(\ell)}, f_2^{(\ell)}) \}_{\ell\in\mathbb{L}}$ with weights $\omega_1$ and $\omega_2$ is given by 
\begin{equation} \label{eq_prod_LMB}
\begin{split}
    & \big(\bm{\pi}_1(\bm{X})\big)^{\omega_1} \big(\bm{\pi}_2(\bm{X})\big)^{\omega_2} \\
    &= \Delta(\bm{X}) \Big( \big[ \tau_1 \big]^{\mathbb{L}-\mathcal{L}(\bm{X})} \big[ \gamma_1 \big]^{\mathcal{L}(\bm{X})} \Big)^{\omega_1} \Big( \big[ \tau_2 \big]^{\mathbb{L}-\mathcal{L}(\bm{X})} \big[ \gamma_2 \big]^{\mathcal{L}(\bm{X})} \Big)^{\omega_2} \\
    & \qquad  \times \Big[\big( f_1 \big)^{\bm{X}} \Big]^{\omega_1} \Big[\big( f_2 \big)^{\bm{X}} \Big]^{\omega_2} \\
    &= \Delta(\bm{X}) \Big[ \big(\tau_1 \big)^{\omega_1} \big(\tau_2 \big)^{\omega_2} \Big]^{\mathbb{L}-\mathcal{L}(\bm{X})} \Big[ \big(\gamma_1 \big)^{\omega_1} \big(\gamma_2 \big)^{\omega_2} \Big]^{\mathcal{L}(\bm{X})} \\
    & \qquad \times \Big[\big( f_1 \big)^{\omega_1} \big( f_2 \big)^{\omega_2}  \Big]^{\bm{X}} \\
    &= \Delta(\bm{X}) \big[ \bar{\tau} \big]^{\mathbb{L}-\mathcal{L}(\bm{X})} \Big[ \big(\gamma_1 \big)^{\omega_1} \big(\gamma_2 \big)^{\omega_2} \Big]^{\mathcal{L}(\bm{X})} \Big[ \eta_f \Big]^{\mathcal{L}(\bm{X})}  \Tilde{f}^{\bm{X}} \\
    &= \Delta(\bm{X}) \big[ \bar{\tau} \big]^{\mathbb{L}-\mathcal{L}(\bm{X})} \Big[ \big(\gamma_1 \big)^{\omega_1} \big(\gamma_2 \big)^{\omega_2} \eta_f \Big]^{\mathcal{L}(\bm{X})} \Tilde{f}^{\bm{X}} \\
    &= \Delta(\bm{X}) \big[ \bar{\tau} \big]^{\mathbb{L}-\mathcal{L}(\bm{X})} \big[\bar{\gamma} \big]^{\mathcal{L}(\bm{X})} \Tilde{f}^{\bm{X}}, 
\end{split}
\end{equation}
where 
\begin{equation}
    \begin{split}
        \bar{\tau} &= \big(\tau_1 \big)^{\omega_1} \big(\tau_2 \big)^{\omega_2} \\
        \bar{\gamma} &= \big(\gamma_1 \big)^{\omega_1} \big(\gamma_2 \big)^{\omega_2} \eta_f.
    \end{split}
\end{equation}

The supremum of the above equation is given by
\begin{equation} 
    \begin{split}
        \sup_{\bm{X}\in\mathbb{X}\times\mathbb{L}} & \big(\bm{\pi}_1(\bm{X})\big)^{\omega_1}  \big(\bm{\pi}_2(\bm{X})\big)^{\omega_2}  \\
        & = \sup_{\bm{X}\in\mathbb{X}\times\mathbb{L}} \Delta(\bm{X}) \big[ \bar{\tau} \big]^{\mathbb{L}-\mathcal{L}(\bm{X})} \big[ \bar{\gamma} \big]^{\mathcal{L}(\bm{X})} \Tilde{f}^{\bm{X}} \\
        & = \max_{L\in\mathbb{L}} \big[ \bar{\tau} \big]^{\mathbb{L}-L} \big[ \bar{\gamma} \big]^{L} \Big[ \sup_{x\in\mathbb{X}} \Tilde{f}(x,\cdot) \Big]^{L} \\
        & = \max_{L\in\mathbb{L}} \big[ \bar{\tau} \big]^{\mathbb{L}-L} \big[ \bar{\gamma} \big]^{L}. 
    \end{split}
\end{equation}
Supposing the above equation reaches maximum when $L= L'$, then we have 
\begin{equation}
    \max_{L\in\mathbb{L}} \big[ \bar{\tau} \big]^{\mathbb{L}-L} \big[ \bar{\gamma} \big]^{L}= \big[ \bar{\tau} \big]^{\mathbb{L}-L'} \big[ \bar{\gamma} \big]^{L'}. 
\end{equation}
Note that for any $\ell\in L'$, $\bar{\gamma} \geq \bar{\tau}$ and vice versa. Therefore, an equivalent form of $\big[ \bar{\tau} \big]^{\mathbb{L}-L'} \big[ \bar{\gamma} \big]^{L'}$ is: $\big[\max\{ \bar{\tau}, \bar{\gamma} \} \big]^{\mathbb{L}}$. The supremum of the two LMB UFS can be simplified to 
\begin{equation} \label{eq_sup_LMB}
    \sup_{\bm{X}\in\mathbb{X}\times\mathbb{L}} \big(\bm{\pi}_1(\bm{X})\big)^{\omega_1} \big(\bm{\pi}_2(\bm{X})\big)^{\omega_2} = \big[\max\{ \bar{\tau}, \bar{\gamma} \} \big]^{\mathbb{L}}. 
\end{equation}

Substituting Eqs.~\eqref{eq_prod_LMB} and \eqref{eq_sup_LMB} into Eq.~\eqref{eq_lmb_fusion} gives us 
\begin{equation}
    \begin{split}
          & \dfrac{\big(\bm{\pi}_1(\bm{X})\big)^{\omega_1} \big(\bm{\pi}_2(\bm{X})\big)^{\omega_2}}{\sup_{\bm{X}\in\mathbb{X}\times\mathbb{L}} \big(\bm{\pi}_1(\bm{X})\big)^{\omega_1} \big(\bm{\pi}_2(\bm{X})\big)^{\omega_2}} \\
          &= \Delta(\bm{X}) \frac{ \big[ \bar{\tau} \big]^{\mathbb{L}-\mathcal{L}(\bm{X})} \big[\bar{\gamma} \big]^{\mathcal{L}(\bm{X})} }{\big[\max\{ \bar{\tau}, \bar{\gamma} \} \big]^{\mathbb{L}}} \Tilde{f}^{\bm{X}} \\
          &= \Delta(\bm{X}) \frac{ \big[ \bar{\tau} \big]^{\mathbb{L}-\mathcal{L}(\bm{X})} \big[\bar{\gamma} \big]^{\mathcal{L}(\bm{X})} }{\big[\max\{ \bar{\tau}, \bar{\gamma} \} \big]^{\mathbb{L}-\mathcal{L}(\bm{X})} \big[\max\{ \bar{\tau}, \bar{\gamma} \} \big]^{\mathcal{L}(\bm{X})} } \Tilde{f}^{\bm{X}} \\
          &= \Delta(\bm{X}) \bigg[ \frac{ \bar{\tau} }{\big[\max\{ \bar{\tau}, \bar{\gamma} \} \big] }\bigg]^{\mathbb{L}-\mathcal{L}(\bm{X})} \bigg[\frac{\bar{\gamma}}{\big[\max\{ \bar{\tau}, \bar{\gamma} \} \big]} \bigg]^{\mathcal{L}(\bm{X})} \Tilde{f}^{\bm{X}} \\
          &= \Delta(\bm{X}) \big[\Tilde{r}_0 \big]^{\mathbb{L}-\mathcal{L}(\bm{X})} \big[\Tilde{r}_1 \big]^{\mathcal{L}(\bm{X})} \Tilde{f}^{\bm{X}}. 
    \end{split}
\end{equation}
This is an LMB with parameters $\Tilde{\bm{\pi}} = \{ ( \Tilde{\tau}^{(\ell)}, \Tilde{\gamma}^{(\ell)}, \Tilde{f}^{(\ell)} ) \}_{\ell\in\mathbb{L}}$.

\end{proof}

% If you have an EPS/PDF photo (graphicx package needed), extra braces are needed around the contents of the optional argument to biography to prevent the LaTeX parser from getting confused when it sees the complicated $\backslash${\tt{includegraphics}} command within an optional argument. (You can create your own custom macro containing the $\backslash${\tt{includegraphics}} command to make things  simpler here.)
 

% \end{IEEEbiography}



\bibliographystyle{IEEEtran}
\bibliography{Ref}

% \vfill

\end{document}